\documentclass[12pt,a4paper]{article}

\DeclareMathSizes{12}{12}{7}{5}
\usepackage{amsfonts}
\usepackage{bm}
\usepackage[dvips]{graphicx}
\usepackage{latexsym,amsmath,amssymb,graphics,stmaryrd}
\usepackage{cite}
\usepackage{array}
\usepackage{subfigure}
\usepackage{multirow}
\usepackage{epsfig}
\usepackage[dvips]{graphicx}
\usepackage[dvips]{color}
\usepackage{pstricks}
\usepackage{pst-node}
\usepackage{pst-plot}
\usepackage{amsthm}
\usepackage{graphics}
\usepackage{subfigure}
\usepackage{fancyhdr}

\usepackage{rotating}
\newlength{\arlength}
\newlength{\arheight}

\usepackage{feynmp}

\usepackage{Dalgebra}

\usepackage{feynmp}

\fancypagestyle{plain}{
\fancyhf{}
\newcommand{\fpage}{\iffloatpage{}{\thepage}}
\fancyfoot[C]{\fpage}

}

\newcommand{\col}{~,}
\newcommand{\pnt}{~.}

\newcommand{\YM}{\text{YM}}

\newcommand{\unitmatrix}{\mathds{1}}

\newcommand{\comm}[2]{\left[#1\smash[b]{\mathbin{,}}#2\right]}

\newcommand{\de}{\operatorname{d}\!}

\newcommand{\e}{\operatorname{e}}

\newlength{\neglength}
\newlength{\diameter}

\newcommand{\svertex}[3][0.5]{%
\fmfiequ{#2}{point #1*length(#3) of #3}
}
\newcommand{\dvertex}[3]{%
\fmfiequ{#1}{point length(#3)/3 of #3}
\fmfiequ{#2}{point 2length(#3)/3 of #3}
}

\newcommand{\chionei}[6][plain]{%
\fmftop{v1}
\fmfbottom{v3}
\fmfforce{(0.125w,h)}{v1}
\fmfforce{(0.125w,0)}{v3}
\fmffixed{(0.25w,0)}{v1,v2}
\fmffixed{(0.25w,0)}{v3,v4}
\fmf{#5,tension=0.5,right=0.25}{v1,vc1}
\fmf{#6,tension=0.5,left=0.25}{v2,vc1}
\fmf{#4,tension=1.25}{vc1,vc2}
\fmf{#2,tension=0.5,left=0.25}{v3,vc2}
\fmf{#3,tension=0.5,right=0.25}{v4,vc2}
\fmf{#1,tension=0.5,right=0,width=1mm}{v3,v4}
\fmfposition
\fmfipath{p[]}
\fmfipair{vd[],vm[],vu[]}
\fmfiset{p1}{vpath(__v1,__vc1)}
\fmfiset{p2}{vpath(__v2,__vc1)}
\fmfiset{p3}{vpath(__vc1,__vc2)}
\fmfiset{p4}{vpath(__v3,__vc2)}
\fmfiset{p5}{vpath(__v4,__vc2)}
\svertex{vm1}{p1}
\dvertex{vu1}{vd1}{p1}
\svertex{vm2}{p2}
\dvertex{vu2}{vd2}{p2}
\svertex{vm3}{p3}
\dvertex{vu3}{vd3}{p3}
\svertex{vm4}{p4}
\dvertex{vd4}{vu4}{p4}
\svertex{vm5}{p5}
\dvertex{vd5}{vu5}{p5}
}

\newcommand{\chionetwo}[1][black]{%
\fmftop{v1}
\fmfbottom{v4}
\fmfforce{(0.125w,h)}{v1}
\fmfforce{(0.125w,0)}{v4}
\fmffixed{(0.25w,0)}{v1,v2}
\fmffixed{(0.25w,0)}{v2,v3}
\fmffixed{(0.25w,0)}{v4,v5}
\fmffixed{(0.25w,0)}{v5,v6}
\fmffixed{(0,whatever)}{vc1,vc3}
\fmffixed{(0,whatever)}{vc2,vc4}
\fmf{plain,tension=0.5,right=0.25}{v1,vc1}
\fmf{plain,tension=0.5,left=0.25}{v2,vc1}
\fmf{phantom,tension=0.5,right=0.25}{v2,vc2}
\fmf{plain,tension=0.5,left=0.25}{v3,vc2}
\fmf{plain,tension=0.5,left=0.25}{v4,vc3}
\fmf{phantom,tension=0.5,right=0.25}{v5,vc3}
\fmf{plain,tension=0.5,left=0.25}{v5,vc4}
\fmf{plain,tension=0.5,right=0.25}{v6,vc4}
\fmf{plain,tension=1.25,left=0}{vc1,vc3}
\fmf{plain,tension=1.25,left=0}{vc2,vc4}
\fmffreeze
\fmf{plain,tension=1,left=0}{vc2,vc3}
\fmf{plain,tension=0.5,right=0,width=1mm}{v4,v6}
\fmffreeze
\fmfposition
\fmfipath{p[]}
\fmfipair{vd[],vm[],vu[]}
\fmfiset{p1}{vpath(__v1,__vc1)}
\fmfiset{p2}{vpath(__v2,__vc1)}
\fmfiset{p6}{vpath(__v3,__vc2)}
\fmfiset{p4}{vpath(__v4,__vc3)}
\fmfiset{p8}{vpath(__v5,__vc4)}
\fmfiset{p9}{vpath(__v6,__vc4)}
\fmfiset{p3}{vpath(__vc1,__vc3)}
\fmfiset{p7}{vpath(__vc2,__vc4)}
\fmfiset{p5}{vpath(__vc2,__vc3)}
\svertex{vm1}{p1}
\svertex{vm2}{p2}
\svertex{vm3}{p3}
\svertex{vm4}{p4}
\svertex{vm5}{p5}
\svertex{vm6}{p6}
\svertex{vm7}{p7}
\svertex{vm8}{p8}
\svertex{vm9}{p9}
}

\newcommand{\chionetwoone}[1][black]{%
\fmftop{v1}
\fmfbottom{v4}
\fmfforce{(0.125w,h)}{v1}
\fmfforce{(0.125w,0)}{v4}
\fmffixed{(0.25w,0)}{v1,v2}
\fmffixed{(0.25w,0)}{v2,v3}
\fmffixed{(0.25w,0)}{v4,v5}
\fmffixed{(0.25w,0)}{v5,v6}
\fmffixed{(0,whatever)}{vc1,vc3}
\fmffixed{(0,whatever)}{vb2,vb4}
\fmffixed{(0,whatever)}{vc1,vb1}
\fmffixed{(0,whatever)}{vc1,vb3}
\fmffixed{(whatever,0)}{vb1,vb2}
\fmffixed{(whatever,0)}{vb3,vb4}
\fmf{plain,tension=0.5,right=0.25}{v1,vc1}
\fmf{plain,tension=0.5,left=0.25}{v2,vc1}
\fmf{phantom,tension=0.5,right=0.25}{v2,vb2}
\fmf{plain,tension=0.5,left=0.25}{v3,vb2}
\fmf{plain,tension=0.5,left=0.25}{v4,vc3}
\fmf{plain,tension=0.5,right=0.25}{v5,vc3}
\fmf{phantom,tension=0.5,left=0.25}{v5,vb4}
\fmf{plain,tension=0.5,right=0.25}{v6,vb4}
\fmf{plain,tension=2,left=0}{vc1,vb1}
\fmf{plain,tension=2,left=0}{vb1,vb3}
\fmf{plain,tension=2,left=0}{vb3,vc3}
\fmf{plain,tension=2,left=0}{vb2,vb4}
\fmffreeze
\fmf{plain,tension=2,left=0}{vb1,vb2}
\fmf{plain,tension=2,left=0}{vb3,vb4}
\fmf{plain,tension=0.5,right=0,width=1mm}{v4,v6}
\fmffreeze
\fmfposition
\fmfipath{p[]}
\fmfipair{vd[],vm[],vu[]}
\fmfiset{p1}{vpath(__v1,__vc1)}
\fmfiset{p2}{vpath(__v2,__vc1)}
\fmfiset{p3}{vpath(__vc1,__vb1)}
\fmfiset{p4}{vpath(__vb1,__vb3)}
\fmfiset{p5}{vpath(__vb1,__vb2)}
\fmfiset{p6}{vpath(__v3,__vb2)}
\fmfiset{p7}{vpath(__vb2,__vb4)}
\fmfiset{p8}{vpath(__vb3,__vb4)}
\fmfiset{p9}{vpath(__v6,__vb4)}
\fmfiset{p10}{vpath(__vb3,__vc3)}
\fmfiset{p11}{vpath(__v4,__vc3)}
\fmfiset{p12}{vpath(__v5,__vc3)}
\svertex{vm1}{p1}
\svertex{vm2}{p2}
\svertex{vm3}{p3}
\svertex{vm4}{p4}
\svertex{vm5}{p5}
\svertex{vm6}{p6}
\svertex{vm7}{p7}
\svertex{vm8}{p8}
\svertex{vm9}{p9}
\svertex{vm10}{p10}
\svertex{vm11}{p11}
\svertex{vm12}{p12}
}

\newcommand{\chionetwothree}[1][black]{%
\fmftop{v1}
\fmfbottom{v5}
\fmfforce{(0.125w,h)}{v1}
\fmfforce{(0.125w,0)}{v5}
\fmffixed{(0.25w,0)}{v1,v2}
\fmffixed{(0.25w,0)}{v2,v3}
\fmffixed{(0.25w,0)}{v3,v4}
\fmffixed{(0.25w,0)}{v5,v6}
\fmffixed{(0.25w,0)}{v6,v7}
\fmffixed{(0.25w,0)}{v7,v8}
\fmffixed{(0,whatever)}{vc1,vc4}
\fmffixed{(0,whatever)}{vc2,vc5}
\fmffixed{(0,whatever)}{vc3,vc6}

\fmf{plain,tension=0.5,right=0.25}{v1,vc1}
\fmf{plain,tension=0.5,left=0.25}{v2,vc1}
\fmf{phantom,tension=0.5,right=0.25}{v2,vc2}
\fmf{plain,tension=0.5,left=0.25}{v3,vc2}
\fmf{phantom,tension=0.5,right=0.25}{v3,vc3}
\fmf{plain,tension=0.5,left=0.25}{v4,vc3}
\fmf{plain,tension=0.5,left=0.25}{v5,vc4}
\fmf{phantom,tension=0.5,right=0.25}{v6,vc4}
\fmf{plain,tension=0.5,left=0.25}{v6,vc5}
\fmf{phantom,tension=0.5,right=0.25}{v7,vc5}
\fmf{plain,tension=0.5,left=0.25}{v7,vc6}
\fmf{plain,tension=0.5,right=0.25}{v8,vc6}
\fmf{plain,tension=1.25,left=0}{vc1,vc4}
\fmf{plain,tension=1.25,left=0}{vc2,vc5}
\fmf{plain,tension=1.25,left=0}{vc3,vc6}
\fmffreeze
\fmf{plain,tension=1,left=0}{vc4,vc2}
\fmf{plain,tension=1,left=0}{vc5,vc3}
\fmf{plain,tension=0.5,right=0,width=1mm}{v5,v8}
\fmffreeze
\fmfposition
}

\newcommand{\chitwoonethree}[1][black]{%
\fmftop{v1}
\fmfbottom{v5}
\fmfforce{(0.125w,h)}{v1}
\fmfforce{(0.125w,0)}{v5}
\fmffixed{(0.25w,0)}{v1,v2}
\fmffixed{(0.25w,0)}{v2,v3}
\fmffixed{(0.25w,0)}{v3,v4}
\fmffixed{(0.25w,0)}{v5,v6}
\fmffixed{(0.25w,0)}{v6,v7}
\fmffixed{(0.25w,0)}{v7,v8}
\fmffixed{(whatever,0.5h)}{v5,vc1}
\fmffixed{(0,whatever)}{vc1,vc4}
\fmffixed{(0,whatever)}{vc2,vc5}
\fmffixed{(0,whatever)}{vc3,vc6}
\fmffixed{(whatever,0)}{vc1,vc3}
\fmffixed{(whatever,0)}{vc3,vc5}

\fmf{plain,tension=0.5,right=0.125}{v1,vc1}
\fmf{phantom,tension=0.5,left=0.25}{v2,vc1}
\fmf{plain,tension=0.5,right=0.25}{v2,vc2}
\fmf{plain,tension=0.5,left=0.25}{v3,vc2}
\fmf{phantom,tension=0.5,right=0.25}{v3,vc3}
\fmf{plain,tension=0.5,left=0.125}{v4,vc3}
\fmf{plain,tension=0.5,left=0.25}{v5,vc4}
\fmf{plain,tension=0.5,right=0.25}{v6,vc4}
\fmf{phantom,tension=0.5,left=0.25}{v6,vc5}
\fmf{phantom,tension=0.5,right=0.25}{v7,vc5}
\fmf{plain,tension=0.5,left=0.25}{v7,vc6}
\fmf{plain,tension=0.5,right=0.25}{v8,vc6}
\fmf{plain,tension=1.25,left=0}{vc1,vc4}
\fmf{plain,tension=1.25,left=0}{vc2,vc5}
\fmf{plain,tension=1.25,left=0}{vc3,vc6}
\fmffreeze
\fmf{plain,tension=1,left=0}{vc1,vc5}
\fmf{plain,tension=1,left=0}{vc5,vc3}
\fmf{plain,tension=0.5,right=0,width=1mm}{v5,v8}
\fmffreeze
\fmfposition
}

\newcommand{\chionethreetwo}[1][black]{%
\fmftop{v1}
\fmfbottom{v5}
\fmfforce{(0.125w,h)}{v1}
\fmfforce{(0.125w,0)}{v5}
\fmffixed{(0.25w,0)}{v1,v2}
\fmffixed{(0.25w,0)}{v2,v3}
\fmffixed{(0.25w,0)}{v3,v4}
\fmffixed{(0.25w,0)}{v5,v6}
\fmffixed{(0.25w,0)}{v6,v7}
\fmffixed{(0.25w,0)}{v7,v8}
\fmffixed{(whatever,0.5h)}{v5,vc2}
\fmffixed{(0,whatever)}{vc1,vc4}
\fmffixed{(0,whatever)}{vc2,vc5}
\fmffixed{(0,whatever)}{vc3,vc6}
\fmffixed{(whatever,0)}{vc2,vc4}
\fmffixed{(whatever,0)}{vc4,vc6}
\fmf{plain,tension=0.5,right=0.25}{v1,vc1}
\fmf{plain,tension=0.5,left=0.25}{v2,vc1}
\fmf{phantom,tension=0.5,right=0.25}{v2,vc2}
\fmf{phantom,tension=0.5,left=0.25}{v3,vc2}
\fmf{plain,tension=0.5,right=0.25}{v3,vc3}
\fmf{plain,tension=0.5,left=0.25}{v4,vc3}
\fmf{plain,tension=0.5,left=0.125}{v5,vc4}
\fmf{phantom,tension=0.5,right=0.25}{v6,vc4}
\fmf{plain,tension=0.5,left=0.25}{v6,vc5}
\fmf{plain,tension=0.5,right=0.25}{v7,vc5}
\fmf{phantom,tension=0.5,left=0.25}{v7,vc6}
\fmf{plain,tension=0.5,right=0.125}{v8,vc6}
\fmf{plain,tension=1.25,left=0}{vc1,vc4}
\fmf{plain,tension=1.25,left=0}{vc2,vc5}
\fmf{plain,tension=1.25,left=0}{vc3,vc6}
\fmffreeze
\fmf{plain,tension=1,left=0}{vc2,vc4}
\fmf{plain,tension=1,left=0}{vc2,vc6}
\fmf{plain,tension=0.5,right=0,width=1mm}{v5,v8}
\fmffreeze
\fmfposition
}

\newcommand{\chionetwothreetwo}[1][black]{%
\fmftop{v1}
\fmfbottom{v5}
\fmfforce{(0.125w,h)}{v1}
\fmfforce{(0.125w,0)}{v5}
\fmffixed{(0.25w,0)}{v1,v2}
\fmffixed{(0.25w,0)}{v2,v3}
\fmffixed{(0.25w,0)}{v3,v4}
\fmffixed{(0.25w,0)}{v5,v6}
\fmffixed{(0.25w,0)}{v6,v7}
\fmffixed{(0.25w,0)}{v7,v8}
\fmffixed{(0,whatever)}{va1,va2}
\fmffixed{(whatever,0)}{va1,vc1}
\fmffixed{(whatever,0)}{va2,vb3}
\fmffixed{(0,whatever)}{vc1,vc3}
\fmffixed{(0,whatever)}{vb2,vb4}
\fmffixed{(0,whatever)}{vc1,vb1}
\fmffixed{(0,whatever)}{vc1,vb3}
\fmffixed{(whatever,0)}{vb1,vb2}
\fmffixed{(whatever,0)}{vb3,vb4}
\fmf{phantom,tension=0.5,right=0.25}{v2,vc1}
\fmf{plain,tension=0.5,left=0.25}{v3,vc1}
\fmf{phantom,tension=0.5,right=0.25}{v3,vb2}
\fmf{plain,tension=0.5,left=0.25}{v4,vb2}
\fmf{plain,tension=0.5,left=0.25}{v6,vc3}
\fmf{plain,tension=0.5,right=0.25}{v7,vc3}
\fmf{phantom,tension=0.5,left=0.25}{v7,vb4}
\fmf{plain,tension=0.5,right=0.25}{v8,vb4}
\fmf{plain,tension=2,left=0}{vc1,vb1}
\fmf{plain,tension=2,left=0}{vb1,vb3}
\fmf{plain,tension=2,left=0}{vb3,vc3}
\fmf{plain,tension=2,left=0}{vb2,vb4}
\fmffreeze
\fmf{plain,tension=0.5,right=0.25}{v1,va1}
\fmf{plain,tension=0.5,left=0.25}{v2,va1}
\fmf{plain,tension=1}{va1,va2}
\fmf{plain,tension=0}{va2,vc1}
\fmf{plain,tension=0.5,left=0.25}{v5,va2}
\fmf{phantom,tension=0.5,right=0.25}{v6,va2}
\fmf{plain,tension=1,left=0}{vb1,vb2}
\fmf{plain,tension=1,left=0}{vb3,vb4}
\fmf{plain,tension=0.5,right=0,width=1mm}{v5,v8}
}

\newcommand{\chitwothreetwoone}[1][black]{%
\fmftop{v1}
\fmfbottom{v5}
\fmfforce{(0.125w,h)}{v1}
\fmfforce{(0.125w,0)}{v5}
\fmffixed{(0.25w,0)}{v1,v2}
\fmffixed{(0.25w,0)}{v2,v3}
\fmffixed{(0.25w,0)}{v3,v4}
\fmffixed{(0.25w,0)}{v5,v6}
\fmffixed{(0.25w,0)}{v6,v7}
\fmffixed{(0.25w,0)}{v7,v8}
\fmffixed{(0,whatever)}{va1,va2}
\fmffixed{(whatever,0)}{va2,vc3}
\fmffixed{(whatever,0)}{va1,vb1}
\fmffixed{(0,whatever)}{vc1,vc3}
\fmffixed{(0,whatever)}{vb2,vb4}
\fmffixed{(0,whatever)}{vc1,vb1}
\fmffixed{(0,whatever)}{vc1,vb3}
\fmffixed{(whatever,0)}{vb1,vb2}
\fmffixed{(whatever,0)}{vb3,vb4}
\fmf{plain,tension=0.5,right=0.25}{v2,vc1}
\fmf{plain,tension=0.5,left=0.25}{v3,vc1}
\fmf{phantom,tension=0.5,right=0.25}{v3,vb2}
\fmf{plain,tension=0.5,left=0.25}{v4,vb2}
\fmf{phantom,tension=0.5,left=0.25}{v6,vc3}
\fmf{plain,tension=0.5,right=0.25}{v7,vc3}
\fmf{phantom,tension=0.5,left=0.25}{v7,vb4}
\fmf{plain,tension=0.5,right=0.25}{v8,vb4}
\fmf{plain,tension=2,left=0}{vc1,vb1}
\fmf{plain,tension=2,left=0}{vb1,vb3}
\fmf{plain,tension=2,left=0}{vb3,vc3}
\fmf{plain,tension=2,left=0}{vb2,vb4}
\fmffreeze
\fmf{plain,tension=0.5,right=0.25}{v1,va1}
\fmf{phantom,tension=0.5,left=0.25}{v2,va1}
\fmf{plain,tension=1}{va1,va2}
\fmf{plain,tension=0}{va1,vc3}
\fmf{plain,tension=0.5,left=0.25}{v5,va2}
\fmf{plain,tension=0.5,right=0.25}{v6,va2}
\fmf{plain,tension=1,left=0}{vb1,vb2}
\fmf{plain,tension=1,left=0}{vb3,vb4}
\fmf{plain,tension=0.5,right=0,width=1mm}{v5,v8}
}

\newcommand{\chitwoonetwothree}[1][black]{%
\fmftop{v1}
\fmfbottom{v5}
\fmfforce{(0.125w,h)}{v1}
\fmfforce{(0.125w,0)}{v5}
\fmffixed{(0.25w,0)}{v1,v2}
\fmffixed{(0.25w,0)}{v2,v3}
\fmffixed{(0.25w,0)}{v3,v4}
\fmffixed{(0.25w,0)}{v5,v6}
\fmffixed{(0.25w,0)}{v6,v7}
\fmffixed{(0.25w,0)}{v7,v8}
\fmffixed{(0,whatever)}{va1,va2}
\fmffixed{(whatever,0)}{va2,vc3}
\fmffixed{(whatever,0)}{va1,vb1}
\fmffixed{(0,whatever)}{vc1,vc3}
\fmffixed{(0,whatever)}{vb2,vb4}
\fmffixed{(0,whatever)}{vc1,vb1}
\fmffixed{(0,whatever)}{vc1,vb3}
\fmffixed{(whatever,0)}{vb1,vb2}
\fmffixed{(whatever,0)}{vb3,vb4}
\fmf{plain,tension=0.5,right=0.25}{v2,vc1}
\fmf{plain,tension=0.5,left=0.25}{v3,vc1}
\fmf{plain,tension=0.5,right=0.25}{v1,vb2}
\fmf{phantom,tension=0.5,left=0.25}{v2,vb2}
\fmf{plain,tension=0.5,left=0.25}{v6,vc3}
\fmf{phantom,tension=0.5,right=0.25}{v7,vc3}
\fmf{plain,tension=0.5,left=0.25}{v5,vb4}
\fmf{phantom,tension=0.5,right=0.25}{v6,vb4}
\fmf{plain,tension=2,left=0}{vc1,vb1}
\fmf{plain,tension=2,left=0}{vb1,vb3}
\fmf{plain,tension=2,left=0}{vb3,vc3}
\fmf{plain,tension=2,left=0}{vb2,vb4}
\fmffreeze
\fmf{phantom,tension=0.5,right=0.25}{v3,va1}
\fmf{plain,tension=0.5,left=0.25}{v4,va1}
\fmf{plain,tension=1}{va1,va2}
\fmf{plain,tension=0}{va1,vc3}
\fmf{plain,tension=0.5,left=0.25}{v7,va2}
\fmf{plain,tension=0.5,right=0.25}{v8,va2}
\fmf{plain,tension=1,left=0}{vb1,vb2}
\fmf{plain,tension=1,left=0}{vb3,vb4}
\fmf{plain,tension=0.5,right=0,width=1mm}{v5,v8}
}

\newcommand{\chithreeonetwoone}[1][black]{%
\fmftop{v1}
\fmfbottom{v5}
\fmfforce{(0.125w,h)}{v1}
\fmfforce{(0.125w,0)}{v5}
\fmffixed{(0.25w,0)}{v1,v2}
\fmffixed{(0.25w,0)}{v2,v3}
\fmffixed{(0.25w,0)}{v3,v4}
\fmffixed{(0.25w,0)}{v5,v6}
\fmffixed{(0.25w,0)}{v6,v7}
\fmffixed{(0.25w,0)}{v7,v8}
\fmffixed{(0,whatever)}{va1,va2}
\fmffixed{(whatever,0)}{va1,vc1}
\fmffixed{(whatever,0)}{va2,vb1}
\fmffixed{(0,whatever)}{vc1,vc3}
\fmffixed{(0,whatever)}{vb2,vb4}
\fmffixed{(0,whatever)}{vc1,vb1}
\fmffixed{(0,whatever)}{vc1,vb3}
\fmffixed{(whatever,0)}{vb1,vb2}
\fmffixed{(whatever,0)}{vb3,vb4}
\fmf{plain,tension=0.5,right=0.25}{v1,vc1}
\fmf{plain,tension=0.5,left=0.25}{v2,vc1}
\fmf{phantom,tension=0.5,right=0.25}{v2,vb2}
\fmf{phantom,tension=0.5,left=0.25}{v3,vb2}
\fmf{plain,tension=0.5,left=0.25}{v5,vc3}
\fmf{plain,tension=0.5,right=0.25}{v6,vc3}
\fmf{phantom,tension=0.5,left=0.125}{v6,vb4}
\fmf{plain,tension=0.5,right=0.125}{v7,vb4}
\fmf{plain,tension=2,left=0}{vc1,vb1}
\fmf{plain,tension=2,left=0}{vb1,vb3}
\fmf{plain,tension=2,left=0}{vb3,vc3}
\fmf{plain,tension=2,left=0}{vb2,vb4}
\fmffreeze
\fmf{plain,tension=0.5,right=0.25}{v3,va1}
\fmf{plain,tension=0.5,left=0.25}{v4,va1}
\fmf{plain,tension=1}{va1,va2}
\fmf{plain,tension=0}{va2,vb2}
\fmf{phantom,tension=0.5,left=0.125}{v7,va2}
\fmf{plain,tension=0.5,right=0.125}{v8,va2}
\fmf{plain,tension=1,left=0}{vb1,vb2}
\fmf{plain,tension=1,left=0}{vb3,vb4}
\fmf{plain,tension=0.5,right=0,width=1mm}{v5,v8}
}

\newcommand{\chionethreetwothree}[1][black]{%
\fmftop{v1}
\fmfbottom{v5}
\fmfforce{(0.125w,h)}{v1}
\fmfforce{(0.125w,0)}{v5}
\fmffixed{(0.25w,0)}{v1,v2}
\fmffixed{(0.25w,0)}{v2,v3}
\fmffixed{(0.25w,0)}{v3,v4}
\fmffixed{(0.25w,0)}{v5,v6}
\fmffixed{(0.25w,0)}{v6,v7}
\fmffixed{(0.25w,0)}{v7,v8}
\fmffixed{(0,whatever)}{va1,va2}
\fmffixed{(whatever,0)}{va1,vc1}
\fmffixed{(whatever,0)}{va2,vb1}
\fmffixed{(0,whatever)}{vc1,vc3}
\fmffixed{(0,whatever)}{vb2,vb4}
\fmffixed{(0,whatever)}{vc1,vb1}
\fmffixed{(0,whatever)}{vc1,vb3}
\fmffixed{(whatever,0)}{vb1,vb2}
\fmffixed{(whatever,0)}{vb3,vb4}
\fmf{plain,tension=0.5,right=0.25}{v3,vc1}
\fmf{plain,tension=0.5,left=0.25}{v4,vc1}
\fmf{phantom,tension=0.5,right=0.25}{v2,vb2}
\fmf{phantom,tension=0.5,left=0.25}{v3,vb2}
\fmf{plain,tension=0.5,left=0.25}{v7,vc3}
\fmf{plain,tension=0.5,right=0.25}{v8,vc3}
\fmf{plain,tension=0.5,left=0.125}{v6,vb4}
\fmf{phantom,tension=0.5,right=0.125}{v7,vb4}
\fmf{plain,tension=2,left=0}{vc1,vb1}
\fmf{plain,tension=2,left=0}{vb1,vb3}
\fmf{plain,tension=2,left=0}{vb3,vc3}
\fmf{plain,tension=2,left=0}{vb2,vb4}
\fmffreeze
\fmf{plain,tension=0.5,right=0.25}{v1,va1}
\fmf{plain,tension=0.5,left=0.25}{v2,va1}
\fmf{plain,tension=1}{va1,va2}
\fmf{plain,tension=0}{va2,vb2}
\fmf{plain,tension=0.5,left=0.125}{v5,va2}
\fmf{phantom,tension=0.5,right=0.125}{v6,va2}
\fmf{plain,tension=1,left=0}{vb1,vb2}
\fmf{plain,tension=1,left=0}{vb3,vb4}
\fmf{plain,tension=0.5,right=0,width=1mm}{v5,v8}
}

\newcommand{\chionetwoonethree}[1][black]{%
\fmftop{v1}
\fmfbottom{v5}
\fmfforce{(0.125w,h)}{v1}
\fmfforce{(0.125w,0)}{v5}
\fmffixed{(0.25w,0)}{v1,v2}
\fmffixed{(0.25w,0)}{v2,v3}
\fmffixed{(0.25w,0)}{v3,v4}
\fmffixed{(0.25w,0)}{v5,v6}
\fmffixed{(0.25w,0)}{v6,v7}
\fmffixed{(0.25w,0)}{v7,v8}
\fmffixed{(0,whatever)}{va1,va2}
\fmffixed{(whatever,0)}{va1,vb3}
\fmffixed{(whatever,0)}{va2,vc3}
\fmffixed{(0,whatever)}{vc1,vc3}
\fmffixed{(0,whatever)}{vb2,vb4}
\fmffixed{(0,whatever)}{vc1,vb1}
\fmffixed{(0,whatever)}{vc1,vb3}
\fmffixed{(whatever,0)}{vb1,vb2}
\fmffixed{(whatever,0)}{vb3,vb4}
\fmf{plain,tension=0.5,right=0.25}{v1,vc1}
\fmf{plain,tension=0.5,left=0.25}{v2,vc1}
\fmf{phantom,tension=0.5,right=0.125}{v2,vb2}
\fmf{plain,tension=0.5,left=0.125}{v3,vb2}
\fmf{plain,tension=0.5,left=0.25}{v5,vc3}
\fmf{plain,tension=0.5,right=0.25}{v6,vc3}
\fmf{phantom,tension=0.5,left=0.25}{v6,vb4}
\fmf{phantom,tension=0.5,right=0.25}{v7,vb4}
\fmf{plain,tension=2,left=0}{vc1,vb1}
\fmf{plain,tension=2,left=0}{vb1,vb3}
\fmf{plain,tension=2,left=0}{vb3,vc3}
\fmf{plain,tension=2,left=0}{vb2,vb4}
\fmffreeze
\fmf{phantom,tension=0.5,right=0.125}{v3,va1}
\fmf{plain,tension=0.5,left=0.125}{v4,va1}
\fmf{plain,tension=1}{va1,va2}
\fmf{plain,tension=0}{va1,vb4}
\fmf{plain,tension=0.5,left=0.25}{v7,va2}
\fmf{plain,tension=0.5,right=0.25}{v8,va2}
\fmf{plain,tension=1,left=0}{vb1,vb2}
\fmf{plain,tension=1,left=0}{vb3,vb4}
\fmf{plain,tension=0.5,right=0,width=1mm}{v5,v8}
}

\newcommand{\chionetwoonetwoone}[1][black]{%
\fmftop{v1}
\fmfbottom{v4}
\fmfforce{(0.125w,h)}{v1}
\fmfforce{(0.125w,0)}{v4}
\fmffixed{(0.25w,0)}{v1,v2}
\fmffixed{(0.25w,0)}{v2,v3}
\fmffixed{(0.25w,0)}{v4,v5}
\fmffixed{(0.25w,0)}{v5,v6}
\fmffixed{(0,whatever)}{vc1,vc3}
\fmffixed{(0,whatever)}{vb2,vb4}
\fmffixed{(0,whatever)}{vc2,vb2}
\fmffixed{(0,whatever)}{vc4,vb4}
\fmffixed{(0,whatever)}{vc1,vb1}
\fmffixed{(0,whatever)}{vc1,vb3}
\fmffixed{(whatever,0)}{vc1,vc2}
\fmffixed{(whatever,0)}{vb1,vb2}
\fmffixed{(whatever,0)}{vb3,vb4}
\fmffixed{(whatever,0)}{vc3,vc4}
\fmf{plain,tension=0.5,right=0.25}{v1,vc1}
\fmf{plain,tension=0.5,left=0.25}{v2,vc1}
\fmf{phantom,tension=0.5,right=0.25}{v2,vc2}
\fmf{plain,tension=0.5,left=0.25}{v3,vc2}
\fmf{plain,tension=0.5,left=0.25}{v4,vc3}
\fmf{phantom,tension=0.5,right=0.25}{v5,vc3}
\fmf{plain,tension=0.5,left=0.25}{v5,vc4}
\fmf{plain,tension=0.5,right=0.25}{v6,vc4}
\fmf{plain,tension=2,left=0}{vc1,vb1}
\fmf{plain,tension=2,left=0}{vb1,vb3}
\fmf{plain,tension=2,left=0}{vb3,vc3}
\fmf{plain,tension=2,left=0}{vc2,vb2}
\fmf{plain,tension=2,left=0}{vb2,vb4}
\fmf{plain,tension=2,left=0}{vb4,vc4}
\fmffreeze
\fmf{plain,tension=2,left=0}{vb1,vc2}
\fmf{plain,tension=2,left=0}{vb3,vb2}
\fmf{plain,tension=2,left=0}{vc3,vb4}
\fmf{plain,tension=0.5,right=0,width=1mm}{v4,v6}
}

\newcommand{\seone}[9][]{%
\settoheight{\eqoff}{$\times$}%
\setlength{\eqoff}{0.5\eqoff}%
\addtolength{\eqoff}{-7.5\unitlength}%
\raisebox{\eqoff}{%
\fmfframe(1,0)(1,0){%
\begin{fmfchar*}(20,15)
\fmfleft{v1}
\fmfright{v2}
\fmffixed{(0.66w,0)}{vc1,vc2}
\fmf{#4}{v1,vc1}
\fmf{#5}{vc2,v2}
\fmf{phantom,left=#2}{vc1,vc2}
\fmf{phantom,left=#3}{vc2,vc1}
\fmffreeze
\fmfposition
\fmfipath{p[]}
\fmfipair{vm[]}
\fmfiset{p1}{vpath(__vc1,__vc2)}
\fmfiset{p2}{vpath(__vc2,__vc1)}
\fmfi{#6}{subpath (0,length(p1)/2) of p1}
\fmfi{#7}{subpath (length(p1)/2,length(p1)) of p1}
\fmfi{#9}{subpath (0,length(p2)/2) of p2}
\fmfi{#8}{subpath (length(p2)/2,length(p2)) of p2}
\svertex{vm1}{p1}
\svertex{vm2}{p2}
{#1}
\end{fmfchar*}}}}

\newcommand{\csefour}[9][]{%
\settoheight{\eqoff}{$\times$}%
\setlength{\eqoff}{0.5\eqoff}%
\addtolength{\eqoff}{-7.5\unitlength}%
\raisebox{\eqoff}{%
\fmfframe(1,0)(1,0){%
\begin{fmfchar*}(20,15)
\fmfleft{v1}
\fmfright{v2}
\fmffixed{(0.66w,0)}{vc1,vc2}
\fmf{#2}{v1,vc1}
\fmf{#3}{vc2,v2}
\fmf{phantom,left=1}{vc1,vc2}
\fmf{phantom,left=1}{vc2,vc1}
\fmffreeze
\fmfposition
\fmfipath{p[]}
\fmfipair{vm[],vl[],vr[]}
\fmfiset{p1}{vpath(__vc1,__vc2)}
\fmfiset{p2}{vpath(__vc2,__vc1)}
\fmfi{#4}{subpath (0,length(p1)/3) of p1}
\fmfi{#5}{subpath (length(p1)/3,2length(p1)/3) of p1}
\fmfi{#6}{subpath (2length(p1)/3,length(p1)) of p1}
\fmfi{#7}{subpath (0,length(p2)/3) of p2}
\fmfi{#8}{subpath (length(p2)/3,2length(p2)/3) of p2}
\fmfi{#9}{subpath (2length(p2)/3,length(p2)) of p2}
\dvertex{vl1}{vr1}{p1}
\dvertex{vr2}{vl2}{p2}
\fmfiset{p3}{vl1--vl2}
\fmfiset{p4}{vr1--vr2}
\svertex{vm1}{p3}
\svertex{vm2}{p4}
{#1}
\end{fmfchar*}}}}

\newcommand{\intsixpo}[2][20]{%
\settoheight{\eqoff}{$\times$}%
\setlength{\eqoff}{0.5\eqoff}%
\addtolength{\eqoff}{-0\unitlength}%
\raisebox{\eqoff}{%
\fmfframe(0,0)(0,-10){%
\begin{fmfchar*}(#1,20)
  \fmfleft{in}
  \fmfright{out1}
\fmf{phantom}{in,v1}
\fmf{phantom}{out,v4}
\fmfforce{(0,0.5h)}{in}
\fmfforce{(h,0.5h)}{out}
\fmfpoly{phantom}{v1,v6,v5,v4,v3,v2}
\fmffixed{(0.05h,0)}{in,v1}
\fmffixed{(0.9h,0)}{v1,v4}
\fmf{phantom}{vc,v1}
\fmf{phantom}{vc,v2}
\fmf{phantom}{vc,v3}
\fmf{phantom}{vc,v4}
\fmf{phantom}{vc,v5}
\fmf{phantom}{vc,v6}
\fmffreeze
{#2}
\end{fmfchar*}}}
}

\newcommand{\intsixp}[2][20]{%
\settoheight{\eqoff}{$\times$}%
\setlength{\eqoff}{0.5\eqoff}%
\addtolength{\eqoff}{-10\unitlength}%
\raisebox{\eqoff}{%
\fmfframe(0,0)(0,0){%
\begin{fmfchar*}(#1,20)
  \fmfleft{in}
  \fmfright{out1}
\fmf{phantom}{in,v1}
\fmf{phantom}{out,v4}
\fmfforce{(0,0.5h)}{in}
\fmfforce{(h,0.5h)}{out}
\fmfpoly{phantom}{v1,v6,v5,v4,v3,v2}
\fmffixed{(0.05h,0)}{in,v1}
\fmffixed{(0.9h,0)}{v1,v4}
\fmf{phantom}{vc,v1}
\fmf{phantom}{vc,v2}
\fmf{phantom}{vc,v3}
\fmf{phantom}{vc,v4}
\fmf{phantom}{vc,v5}
\fmf{phantom}{vc,v6}
\fmffreeze
{#2}
\end{fmfchar*}}}
}

\newcommand{\fourlint}[9][]{%
\settoheight{\eqoff}{$\times$}%
\setlength{\eqoff}{0.5\eqoff}%
\addtolength{\eqoff}{-10\unitlength}%
\raisebox{\eqoff}{%
\fmfframe(0,0)(0,0){%
\begin{fmfchar*}(20,20)
  \fmfleft{in}
  \fmfright{out1}
\fmf{phantom}{in,v1}
\fmf{phantom}{out,v3}
\fmfforce{(0,0.5h)}{in}
\fmfforce{(w,0.5h)}{out}
\fmfpoly{phantom}{v1,v4,v3,v2}
\fmffixed{(0.9w,0)}{v1,v3}
\fmffixed{(0.05w,0)}{in,v1}
\fmf{#2}{v1,v2}
\fmf{#3}{v2,v3}
\fmf{#4}{v3,v4}
\fmf{#5}{v4,v1}
\fmf{phantom}{vc,v1}
\fmf{phantom}{vc,v3}
\fmffreeze
\fmf{#6}{vc,v1}
\fmf{#7}{vc,v2}
\fmf{#8}{vc,v3}
\fmf{#9}{vc,v4}
{#1}
\end{fmfchar*}}}
}

\DeclareMathOperator{\tr}{tr}

\DeclareMathOperator{\dperm}{\mathbf{P}}
\DeclareMathOperator{\depsilon}{\bm{\epsilon}}
\DeclareMathOperator{\ds}{d}

\DeclareMathOperator{\Fop}{F}
\DeclareMathOperator{\chiop}{\mathbf{\chi}}
\DeclareMathOperator{\D}{D}

\newlength{\eqoff}

\newlength{\unit}
\psset{xunit=\unit,yunit=\unit,runit=\unit}
\newlength{\linew}
\psset{linewidth=\linew}

\usepackage{dsfont}

\numberwithin{equation}{section}

\newcommand{\NN}{{\mathcal N}}
\newcommand{\DD}{{\mathcal D}}
\newcommand{\la}{\lambda}
\begin{document}
\begin{fmffile}{graphs}
\fmfcmd{%
input Dalgebra
}

\fmfcmd{%
def getmid(suffix p) =
  pair p.mid[], p.off[], p.dir[];
  for i=0 upto 36:
    p.dir[i] = dir(5*i);
    p.mid[i]+p.off[i] = directionpoint p.dir[i] of p;
    p.mid[i]-p.off[i] = directionpoint -p.dir[i] of p;
  endfor
enddef;
}

\fmfcmd{%
marksize=2mm;
def draw_mark(expr p,a) =
  begingroup
    save t,tip,dma,dmb; pair tip,dma,dmb;
    t=arctime a of p;
    tip =marksize*unitvector direction t of p;
    dma =marksize*unitvector direction t of p rotated -45;
    dmb =marksize*unitvector direction t of p rotated 45;
    linejoin:=beveled;
    draw (-.5dma.. .5tip-- -.5dmb) shifted point t of p;
  endgroup
enddef;
style_def derplain expr p =
    save amid;
    amid=.5*arclength p;
    draw_mark(p, amid);
    draw p;
enddef;
style_def derphoton expr p =
    save amid;
    amid=.5*arclength p;
    draw_mark(p, amid);
    draw wiggly p;
enddef;
def draw_marks(expr p,a) =
  begingroup
    save t,tip,dma,dmb,dmo; pair tip,dma,dmb,dmo;
    t=arctime a of p;
    tip =marksize*unitvector direction t of p;
    dma =marksize*unitvector direction t of p rotated -45;
    dmb =marksize*unitvector direction t of p rotated 45;
    dmo =marksize*unitvector direction t of p rotated 90;
    linejoin:=beveled;
    draw (-.5dma.. .5tip-- -.5dmb) shifted point t of p withcolor 0white;
    draw (-.5dmo.. .5dmo) shifted point t of p;
  endgroup
enddef;
style_def derplains expr p =
    save amid;
    amid=.5*arclength p;
    draw_marks(p, amid);
    draw p;
enddef;
def draw_markss(expr p,a) =
  begingroup
    save t,tip,dma,dmb,dmo; pair tip,dma,dmb,dmo;
    t=arctime a of p;
    tip =marksize*unitvector direction t of p;
    dma =marksize*unitvector direction t of p rotated -45;
    dmb =marksize*unitvector direction t of p rotated 45;
    dmo =marksize*unitvector direction t of p rotated 90;
    linejoin:=beveled;
    draw (-.5dma.. .5tip-- -.5dmb) shifted point t of p withcolor 0white;
    draw (-.5dmo.. .5dmo) shifted point arctime a+0.25 mm of p of p;
    draw (-.5dmo.. .5dmo) shifted point arctime a-0.25 mm of p of p;
  endgroup
enddef;
style_def derplainss expr p =
    save amid;
    amid=.5*arclength p;
    draw_markss(p, amid);
    draw p;
enddef;
style_def dblderplains expr p =
    save amidm;
    save amidp;
    amidm=.5*arclength p-0.75mm;
    amidp=.5*arclength p+0.75mm;
    draw_mark(p, amidm);
    draw_marks(p, amidp);
    draw p;
enddef;
style_def dblderplainss expr p =
    save amidm;
    save amidp;
    amidm=.5*arclength p-0.75mm;
    amidp=.5*arclength p+0.75mm;
    draw_mark(p, amidm);
    draw_markss(p, amidp);
    draw p;
enddef;
style_def dblderplainsss expr p =
    save amidm;
    save amidp;
    amidm=.5*arclength p-0.75mm;
    amidp=.5*arclength p+0.75mm;
    draw_marks(p, amidm);
    draw_markss(p, amidp);
    draw p;
enddef;
}

%
%

\fmfcmd{%
thin := 1pt; 
thick := 2thin;
arrow_len := 4mm;
arrow_ang := 15;
curly_len := 3mm;
dash_len := 1.5mm; 
dot_len := 1mm; 
wiggly_len := 2mm; 
wiggly_slope := 60;
zigzag_len := 2mm;
zigzag_width := 2thick;
decor_size := 5mm;
dot_size := 2thick;
}


\begingroup\parindent0pt
\vspace*{2em}
\begin{flushright}\footnotesize
\texttt{UUITP-34/11} \\
\texttt{HU-EP-11/62} \\
\texttt{HU-MATH-2011-26}
\vspace{0.8cm}\end{flushright}

\begingroup\LARGE
Four-loop anomalous dimensions in 
Leigh-Strassler
deformations
\par\endgroup
\vspace{1.5em}
\begingroup\large
J.\ A.\ Minahan ${}^a$, C.\ Sieg ${}^b$
\par\endgroup
\vspace{1em}
\begingroup\itshape
$^a$  Department of Physics and Astronomy, Uppsala University\\
SE-751 08 Uppsala, Sweden\\
$^b$ 
Institut f\"ur Mathematik und Institut f\"ur Physik, Humboldt-Universit\"at zu Berlin\\
Johann von Neumann Haus, Rudower Chaussee 25, 12489 Berlin, Germany
\par\endgroup
\vspace{1em}
\begingroup\ttfamily
joseph.minahan@fysast.uu.se\\
csieg@math.hu-berlin.de
\par\endgroup
\vspace{1.5em}
\endgroup

\paragraph{Abstract.}

We determine the scalar part of the four-loop  chiral dilatation operator for  Leigh-Strassler deformations of  $\mathcal{N}=4$ super Yang-Mills.
This is sufficient   to find the four-loop anomalous dimensions for operators in closed scalar subsectors.
This includes the $SU(2)$ subsector of the (complex) $\beta$-deformation, where we explicitly compute the anomalous dimension for operators with a single impurity.  It also includes the ``3-string null" operators of the cubic Leigh-Strassler deformation. 
Our four-loop results show
that the rational part of the anomalous dimension
is consistent with a conjecture made in arXiv:1108.1583 based 
on the three-loop result of arXiv:1008.3351 and the $\NN=4$ 
magnon dispersion relation. 
Here we find additional $\zeta(3)$ terms.




\newpage


\section{Introduction}

\label{sec:introduction}

The spectrum of single trace operators in planar  $\NN=4$ supersymmetric Yang-Mills (SYM) is solvable,  at least  in principle if not always in practice, 
because of an underlying integrability
(see \cite{Beisert:2010jr} for a comprehensive review).
Starting with $\NN=4$ SYM,  there exists a class of deformations  that break
 the superconformal symmetry down to $\NN=1$ but preserve the integrability.  However, there exist other deformations that destroy the 
 integrability and we can ask ourselves to what extent  one can calculate the spectra of these theories.  
  A possible way forward is to compute the   dilatation operator to high enough loop order such that a pattern emerges.   


The general class of deformations of $\NN=4$ SYM that preserve an
$\mathcal{N}=1$ superconformal symmetry were first catalogued 
by 
Leigh and Strassler \cite{Leigh:1995ep}. The Leigh-Strassler 
superpotential is given by
\begin{equation}\label{superpot}
W=i\kappa\Big[\tr(XYZ-q XZY)+\frac{h}{3}\tr(X^3+Y^3+Z^3)\Big]\col
\end{equation}
which 
depends on three complex parameters $\kappa$, $q$ and $h$, although
the imaginary part of $\kappa$ can 
be eliminated
by a chiral field phase rotation.  The deformation is exactly marginal 
if the Yang-Mills 
coupling $g_\YM$ and  the deformation parameters  satisfy the relation
\begin{equation}\label{couprel}
2g_\YM^2=\kappa\bar\kappa(1+q\bar q+h\bar h)+\mathcal{O}((\kappa\bar\kappa)^4)\col
\end{equation}
where we have allowed for a fourth-order correction, to be discussed in 
more detail below.  
This leaves a  three complex dimensional space of 
$\mathcal{N}=1$ superconformal theories.
Setting the coefficients  to $\kappa=g_\YM$, $q=1$, $h=0$, we recover
 the  $\mathcal{N}=4$  superpotential
  \begin{equation}\label{N4superpot}
W=ig_\YM\tr\comm{X\,}{\,Y}Z\pnt
\end{equation}

The deformation \eqref{superpot} includes some  interesting
special cases.
%
The most well known deformation is the so-called $\beta$-deformation
(see \cite{Zoubos:2010kh} for a review)
where $q=\e^{-2i\pi\beta}$, $h=0$ with  $\beta$ real.  Inserting this into 
\eqref{couprel} then yields $\kappa=g_\YM$ which is exact in the planar 
limit \cite{Khoze:2005nd,Mauri:2005pa}.
With a field redefinition, the $\beta$-deformed superpotential  can be 
recast 
into the form
\begin{equation}
W=ig_\YM\tr(e^{ i\pi \beta}XYZ-e^{- i\pi\beta}XZY)\pnt
\end{equation}
Furthermore, the computation 
of  
the spectrum 
for 
local operators is an integrable problem
\cite{Roiban:2003dw,Berenstein:2004ys,Beisert:2005if,Frolov:2005ty}.
In \cite{Berenstein:2004ys} it was shown 
that at the one-loop level 
the corresponding Bethe equations are the same as in 
$\mathcal{N}=4$ SYM,
except for a $\beta$-dependent shift.  This was extended to all loops
in \cite{Beisert:2005if}.  The supergravity dual for the
$\beta$-deformed theory is known \cite{Lunin:2005jy} and its
world-sheet theory has been shown to be classically integrable, even
though it is not a coset \cite{Frolov:2005dj,Frolov:2005ty}. 
Other integrable deformations with nonzero values of $h$ can be obtained by acting on the $\beta$-deformed theories with similarity transformations  \cite{Bundzik:2005zg,Mansson:2007sh,Mansson:2008xv}.

If the deformation is generalized 
to complex $\beta$, 
the relation of the couplings has to be  altered at four-loop order
\cite{Elmetti:2006gr,Elmetti:2007up},
as 
indicated 
 in \eqref{couprel}. 
The resulting
spin-chain is a known integrable model only 
in a certain subsector
\cite{Berenstein:2004ys}.
 This 
subsector plays 
an 
important role in 
$\mathcal{N}=4$ SYM
and in its (complex) $\beta$-deformation, where it is closed, 
at least perturbatively \cite{Minahan:2005jq}.
It consists of operators 
composed of 
two flavors of complex scalar fields. 
The subsector 
is called the $SU(2)$ subsector, since an
$SU(2)$ subgroup of the
$SU(4)$ $R$-symmetry of $\mathcal{N}=4$
transforms the two flavors into 
each other. We will stick with 
this name, even if the $SU(2)$ $R$-symmetry is 
broken in the presence of the (complex) 
$\beta$-deformation.
In a formulation with manifest $\mathcal{N}=1$ supersymmetry, these
complex scalars are the lowest components of chiral superfields.
Hence, being composites of two of these three superfield flavors, the operators of the $SU(2)$ subsector are themselves chiral
superfields. 

There exist a bigger subsector that consists of all possible
chiral composite operators. This means,  the latter can contain 
all three types of chiral superfield flavors $X$, $Y$, $Z$, and also the 
chiral superfield strength $W_\alpha$. 
Starting at two loops, three chiral
field flavors can  transform into two  $W_\alpha$ and {\it vice versa}.
If $\kappa\ne0$, $h=0$ in \eqref{superpot}, this mixing 
can only occur if all three interacting chiral field flavors are different, 
while at  $h\ne0$ a mixing occurs even if all these flavors are identical.
This guarantees that the $SU(2)$ subsector is closed whenever $h=0$, but it 
also means that in the case $h\ne0$ the mixing extends to the bigger subsector.
When restricting to the lowest lying components of the superfields, 
the field content of this bigger subsector is reduced to the three 
complexified scalars and the two-component gaugino $\psi_\alpha$. In $\NN=4$ SYM this is the $SU(2|3)$ 
subsector \cite{Beisert:2003ys}.  


Let us now consider those deformations that have closed scalar subsectors and restrict ourselves to the planar limit. For  the (complex) $\beta$-deformations, the simplest 
single-trace operators 
are 
$L$ copies of 
 a single superfield flavor $X$,
\begin{equation}\label{SU2groundstate}
\tr X\dots X \col
\end{equation}
which 
are protected from quantum corrections. In the spin chain picture
each of these operators 
is 
the ferromagnetic ground state 
of 
 a closed spin chain with fixed length $L$. 
One constructs excited states by changing the 
flavor of 
one or more 
of the 
fields.
The composite operator that contains a single 
changed flavor 
\begin{equation}\label{SU2oneexstate}
\tr YX\dots X 
\end{equation}
corresponds to a spin chain with one excitation
(magnon). 
While in the $\mathcal{N}=4$  case 
this operator is still protected,
in the  (complex) $\beta$-deformed theory it acquires an anomalous dimension.  We can add more excitations by adding more $Y$ fields.  Adding $Z$ 
fields takes us out of the $SU(2)$ subsector.


The only other way to have a closed scalar subsector is to
 take the 
 limit $\kappa\to0$ with fixed $\kappa|h|=\sqrt{2}g_\YM$, such that
the superpotential becomes 
\begin{equation}
W=i\sqrt{2}g_\YM\tr(X^3+Y^3+Z^3)\pnt
\end{equation}
This special case, called the (Fermat) cubic Leigh-Strassler deformation, was recently  analyzed in \cite{Minahan:2011dd}.
This  deformation has no closed subsector resembling the 
$SU(2)$ subsector of the undeformed theory.  However, because the superpotential does not have the ``hopping" term there is no direct mixing of scalar operators in the planar limit.  Scalar mixing can still occur indirectly through intermediate scalar-fermion mixing, but if the operator does not have a sequence of three or more identical scalar fields, then this mixing will  not occur either.  We will call such operators ``3-string null''. 
In the planar limit these operators do not mix with other operators.

In the cubic Leigh-Strassler theory, the ground states are anti-ferromagnetic in nature, having the general form \cite{Minahan:2011dd}
\begin{equation}\label{cubicLSgroundstate}
\tr XYXZYZ\dots XY \col
\end{equation}
where no two neighboring fields have the same flavor.
These operators are
protected 
in the planar limit
to all orders in perturbation theory.  
Operators of the type
\begin{equation}\label{cubiconeexstate}
\tr  XYXYYZ\dots XY
\col 
\end{equation}
where  there are two neighboring fields with the same flavor
correspond to the first excited states. They are not protected,
but they are  3-string null and hence 
do not undergo operator mixing either.
The lack of mixing makes it possible to formulate an all-loop conjecture about the 
rational part of their anomalous dimensions \cite{Minahan:2011dd}.  The conjecture is applicable to any 3-string null operator \cite{Minahan:2011dd}.

The upshot is that we have two classes of Leigh-Strassler theories where the mixing is closed to scalar operators; the $SU(2)$ subsector of the complex $\beta$-deformed theories and the 3-string null operators in the cubic Leigh-Strassler theory.  For these two cases we only need the scalar part of 
the ``chiral'' dilatation operator, where 
``chiral'' indicates the restriction to the field content of the
$SU(2|3)$ subsector.



In this paper we determine  the scalar part of the chiral
dilatation operator to four-loop order.   
The evaluation of the relevant Feynman diagrams becomes manageable by 
determining the deviations from the four-loop dilatation operator in the $SU(2)$ subsector of the $\NN=4$ SYM theory. The latter has been determined \cite{Beisert:2007hz} 
 as one of 
the conserved local charges \cite{Beisert:2003tq,Beisert:2005wv} using integrability and the postulated form of the magnon dispersion relation.
Hence, by doing perturbation theory in the general Leigh-Strassler case, we 
determine the scalar part of the chiral dilatation operator. These more general deformations are not believed to be integrable.
But our result is also valid for the mixing between all three 
scalar chiral fields of $\mathcal{N}=4$ SYM.


 The scalar part we find for the chiral dilatation operator is valid for {\it any} Leigh-Strassler deformation.   It is complete when acting on a closed scalar subsector, hence we can apply it to the above cases to find the anomalous dimensions to four-loop order.  When acting on the 3-string null operators in the cubic Leigh-Strassler theory we find that the rational contributions to the anomalous dimensions are consistent with an all-loop conjecture made in  \cite{Minahan:2011dd}.  Furthermore, we see that the  four-loop  anomalous dimensions contain  additional  transcendental terms.


The paper is organized as follows.
In section  \ref{sec:Feynmandiag} we analyze the Feynman diagrams that
modify the scalar part of the chiral four-loop dilatation operator 
of $\mathcal{N}=4$ SYM theory
for  Leigh-Strassler deformations.
In section \ref{sec:oneimpstates} we simplify the results from the previous 
section and
 determine the  anomalous dimensions of
the one impurity  operators in the 
(complex) $\beta$-deformation and the 3-string null operators in the cubic Leigh-Strassler deformation.
In section \ref{sec:conclusions} we draw our conclusions. 
Several technical details concerning the so-called chiral functions, 
similarity transformations and relevant loop integrals
are included in various appendices.

\section{Feynman diagram analysis}
\label{sec:Feynmandiag}

The perturbative spectrum of local composite operators in a conformal field theory is given by the bare dimensions of the operators plus their anomalous dimensions.
The latter are generated 
by the renormalization of the composite operators 
\begin{equation}\label{opren}
\mathcal{O}_{a,\text{ren}}
=\mathcal{Z}_{a}{}^b(\lambda,\varepsilon)\mathcal{O}_{b,\text{bare}}
\col
\end{equation}
which in general imply a mixing among the operators.
$\mathcal{Z}$ is the matrix-valued renormalization constant
that is given as a power series in the 
 't Hooft coupling constant 
$\lambda=g_\YM^2N$ and 
absorbs the overall UV divergences
which are 
generated at each loop order. 
In $D=4-2\varepsilon$ dimensions the divergences appear 
as poles in $\varepsilon$.

We determine the anomalous dimensions as eigenvalues of 
the dilatation operator, which in terms of $\mathcal{Z}$ is defined as
\begin{equation}\label{DinZ}
\mathcal{D}
=\mu\frac{\de}{\de\mu}\ln\mathcal{Z}(\lambda\mu^{2\varepsilon},\varepsilon)
=\lim_{\varepsilon\rightarrow0}\left[2\varepsilon\lambda
\frac{\de}{\de \lambda}\ln\mathcal{Z}(\lambda,\varepsilon)\right]
\pnt
\end{equation}
The logarithm in the above description is understood as a series 
expansion in
the orthonormalized basis of composite operators 
for the unrenormalized theory, such that the leading contribution
to $\mathcal{Z}$ is the identity.
All higher order poles in $\ln\mathcal{Z}$ must 
 cancel, leading to 
 the second relation in \eqref{DinZ}. 
The anomalous dimensions are then the eigenvalues of $\mathcal{D}$.

We first calculate the renormalization constant $\mathcal{Z}$
for operators composed of scalar chiral superfields
by computing the relevant Feynman diagrams 
using an $\mathcal{N}=1$ superfield formulation.
We then
use \eqref{DinZ} to derive 
the dilatation operator.
For a more detailed description 
we refer the reader 
to \cite{Sieg:2010jt,Sieg:2010tz}. It should be understood that $\DD$ is the scalar part of the chiral dilatation operator, since 
we have restricted ourselves to scalar chiral operators. 
The dilatation operator can be expressed as the series expansion
\begin{eqnarray}
\DD=\sum_{n=1}^\infty g^{2n}\DD_n\col
\qquad
g\equiv\frac{\sqrt{\la}}{4\pi}\pnt
\end{eqnarray}
For $\mathcal{N}=4$ SYM, the first three terms in the expansion were found to be 
\begin{equation}\label{D1D2D3}
\begin{aligned}
\mathcal{D}_1
&={}-{}2\chi(1)
\col\\
\mathcal{D}_2
&={}-{}2[\chi(1,2)+\chi(2,1)]+4\chi(1)
\col\\
\mathcal{D}_3
&=-4(\chi(1,2,3)+\chi(3,2,1))+4i\epsilon_2[\chi(2,1,3)-\chi(1,3,2)]-4\chi(1,3)\\
&\phantom{{}={}}
+16(\chi(1,2)+\chi(2,1))-16\chi(1)-4(\chi(1,2,1)+\chi(2,1,2))
\col
\end{aligned}
\end{equation}
where $\epsilon_2=-\frac{i}{2}$ is a parameter that can be changed by
similarity transformations of the basis of operators.

The terms in \eqref{D1D2D3} are expressed using the very convenient 
basis of so-called chiral functions. 
They capture the structure of the chiral and
anti-chiral superfields within the Feynman diagrams 
\cite{Fiamberti:2007rj,Fiamberti:2008sh} 
(see \cite{Sieg:2010jt} for a review).
Here they are defined as
\begin{equation}\label{chifuncdef}
\begin{aligned}
\chiop(a_1,\dots,a_n)=\sum_{r=0}^{L-1}\prod_{i=1}^n
\Fop_{a_i+r\;a_i+r+1}
\pnt
\end{aligned}
\end{equation}
where the operator $\Fop_{ij}$ acts on the flavors at sites $i$ and $j$.
Its precise form will be given below.  
The product of the $\Fop_{ij}$ is specified
by the argument list of the chiral function. 
Periodicity in $L$ is understood 
when this product acts on adjacent elementary flavors of a single-trace 
operator with the length $L$, i.e. with $L$ elementary fields. 
Shifts of all integer entries $a_i$, 
$i=1,\dots,n$ in the argument lists hence do not produce new chiral functions.
We can therefore choose $\min(a_1,\dots,a_n)=1$.
 
In the $\mathcal{N}=4$ SYM theory, the operator $\Fop_{ij}$
is given as the permutation minus the identity. 
In fact, if we consider the Leigh-Strassler 
deformation, only the chiral functions
\eqref{chifuncdef} are modified since $\Fop_{ij}$ 
as a contraction of the 
superpotential \eqref{superpot} with its Hermitean conjugate
is sensitive to the deformation.
There is no further modification of the expressions in 
\eqref{D1D2D3} since they rely on a full Feynman diagram calculation in 
\cite{Sieg:2010tz} that does not make use of relations
between chiral functions.
With the superpotential \eqref{superpot} the fundamental building block
$\Fop_{ij}$ evaluates to
\begin{equation}\label{buildingblock}
\begin{aligned}
\settoheight{\eqoff}{$\times$}%
\setlength{\eqoff}{0.5\eqoff}%
\addtolength{\eqoff}{-13\unitlength}%
\raisebox{\eqoff}{%
\fmfframe(-1.5,3)(-11.5,3){%
\begin{fmfchar*}(20,20)
\chionei[phantom]{plain}{plain}{plain}{plain}{plain}
\fmfiv{label=$\scriptstyle i$,l.a=-90,l.dist=4}{vloc(__v3)}
\fmfiv{label=$\scriptstyle j$,l.a=-90,l.dist=4}{vloc(__v4)}
\end{fmfchar*}}}
&=\Fop_{ij}
=\rho^2(\dperm-\unitmatrix+
\bar h\depsilon \ds-h\ds\bar\depsilon+h\bar h\ds\ds)_{ij}
\col\quad
\rho=\frac{\kappa}{g_\YM}=\sqrt{\frac{2}{1+q\bar q+h\bar h}}+\mathcal{O}(\kappa^6)
\col
\end{aligned}
\end{equation}
where we have used the relation of the couplings \eqref{couprel} and 
the aforementioned field redefinitions in order to make $\kappa$ real. 
The permutation $\dperm$ and $\depsilon$ are
generalizations of the ordinary permutation and the 
$\epsilon$-tensor, which take into account 
the deformation by $q$. 
The only non-vanishing components of the tensors in the products 
up to cyclic permutations are
\begin{equation}
\begin{gathered}
\depsilon_{XYZ}=1\col\quad\depsilon_{YXZ}=-q\col\quad
\bm{\bar\epsilon}_{XYZ}=-\bar q\col\quad\bm{\bar\epsilon}_{YXZ}=1
\col\quad
d_{XXX}=d_{YYY}=d_{ZZZ}=1\pnt
\end{gathered}
\end{equation}
Using $i_i$, $i_j$ and $o_i$, $o_j$ as the flavor degrees of freedom
that enter and leave the building block at the respective positions $i$ and $j$,
the individual terms in \eqref{buildingblock} read
\begin{equation}
\begin{aligned}[]
[(\dperm-\unitmatrix)_{ij}]_{i_ii_j}^{o_io_j}&=-\bm{\epsilon}^{o_io_jk}\bar{\bm{\epsilon}}_{ki_ii_j}\col\\
[(\depsilon \ds)_{ij}]_{i_ii_j}^{o_io_j}&=\bm{\epsilon}^{o_io_jk}d_{ki_ii_j}\col\\
[(\ds\bar\depsilon)_{ij}]_{i_ii_j}^{o_io_j}&=d^{o_io_jk}\bar{\bm{\epsilon}}_{ki_ii_j}\col\\
[(\ds\bar\ds)_{ij}]_{i_ii_j}^{o_io_j}&=d^{o_io_jk}d_{ki_ii_j}
\pnt
\end{aligned}
\end{equation}
The above combinations can be written in terms of bilinears in the 
Gell-Mann matrices $\lambda^a$, $a=1,\dots,8$, which act on the 
flavors at positions $i$ and $j$.  
We form  combinations
$\lambda^{\pm3}=\frac{1}{2}(\lambda^1\pm i\lambda^2)$,
$\lambda^{\pm2}=\frac{1}{2}(\lambda^4\pm i\lambda^5)$,
$\lambda^{\pm1}=\frac{1}{2}(\lambda^6\pm i\lambda^7)$,
such that $\lambda^{\pm i}$ act as rising and lowering 
operators, leaving unaffected the flavor of type $i$, and also 
introduce the identity $\lambda^0=\sqrt{\frac{2}{3}}\unitmatrix$, 
such that $\tr\lambda^A\lambda^B=2\delta^{AB}$, $A,B=0,1,\dots 8$.
In terms of these matrices, the expressions read
\begin{equation}\label{permdef}
\begin{aligned}
(\dperm-\unitmatrix)_{ij}&=\frac{1}{4}
(1+q\bar q)\big(\lambda_i^3\lambda_j^3+\lambda_i^8\lambda_j^8
-2\lambda_i^0\lambda_j^0\big)
+\frac{\sqrt{3}}{4}(1-q\bar q)\big(\lambda_i^8\lambda_j^3-\lambda_i^3\lambda_j^8\big)\\
&\phantom{{}={}}
+q\big(\lambda_i^{-1}\lambda_j^{+1}+\lambda_i^{+2}\lambda_j^{-2}+\lambda_i^{-3}\lambda_j^{+3}\big)
+\bar q\big(\lambda_i^{+1}\lambda_j^{-1}+\lambda_i^{-2}\lambda_j^{+2}+\lambda_i^{+3}\lambda_j^{-3}\big)
\col\\
(\depsilon \ds)_{ij}&=\lambda_i^{+1}\lambda_j^{-3}+\lambda_i^{-2}\lambda_j^{-1}+\lambda_i^{+3}\lambda_j^{+2}
-\bar q\big(\lambda_i^{-1}\lambda_j^{-2}+\lambda_i^{+2}\lambda_j^{+3}+\lambda_i^{-3}\lambda_j^{+1}\big)\col\\
(\ds \depsilon)_{ij}&=\lambda_i^{-1}\lambda_j^{+3}+\lambda_i^{+2}\lambda_j^{+1}+\lambda_i^{-3}\lambda_j^{-2}
-q\big(\lambda_i^{+1}\lambda_j^{+2}+\lambda_i^{-2}\lambda_j^{-3}+\lambda_i^{+3}\lambda_j^{-1}\big)\col\\
\ds\ds_{ij}&=\frac{1}{2}\big(\lambda_i^3\lambda_j^3+\lambda_i^8\lambda_j^8+\lambda_i^0\lambda_j^0\big)
\pnt
\end{aligned}
\end{equation}

A projection of the above expressions onto the subspace of operators that 
contain only the flavors of type $X$ and $Y$, is realized as 
\begin{equation}\label{lambdaSU2proj}
\lambda^{\pm3}\to\sigma^{\pm}\col\qquad
\lambda^{3}\to\sigma^{3}\col\qquad
\lambda^{8}\to\frac{1}{\sqrt{3}}\unitmatrix
\col\qquad
\lambda^{0}\to\sqrt{\frac{2}{3}}\unitmatrix
\col
\end{equation}
where $\sigma^{\pm}$, $\sigma^3$, $\unitmatrix$ are the
Pauli and identity matrices in the two-dimensional flavor space. 
Moreover, we have to drop all combinations of $\lambda^{\pm i}\lambda^{\pm i}$
with $i=1,2$.
The combination $\dperm-\unitmatrix$ then reduces to the respective
expression in the $SU(2)$ subsector, and also $\ds\ds$ can be consistently 
truncated, while the terms linear in 
$h$, $\bar h$ in \eqref{buildingblock} do not admit a restriction to 
two types of flavors.

\subsection{Self energy diagrams}

In a superconformal theory, the anomalous dimensions of the chiral superfields must vanish.  This imposes the relation on the couplings in  \eqref{couprel}.
At one loop, the vanishing of the anomalous dimensions 
is equivalent to the UV finiteness of the 
theory. 
Examining the one-loop diagrams
\begin{equation}\label{sediag}
\begin{aligned}
\seone{1}{1}{plain}{plain}{plain}{plain}{plain}{plain}
=N\kappa\bar\kappa(1+q\bar q+h\bar h)I_1
\col\qquad
\seone{1}{0}{plain}{plain}{photon}{photon}{plain}{plain}
=-2Ng_\YM^2I_1
\col
\end{aligned}
\end{equation}
one can immediately see that the cancellation of the divergences 
leads to the relation in \eqref{couprel}
at leading order. 
This continues to hold  
at two- and three-loop order, where the one-loop  
relation \eqref{couprel}
is not altered 
\cite{Elmetti:2006gr,Elmetti:2007up,Bork:2007bj}.
The reason is that the chiral field lines in the 
two-point 
Feynman diagrams
only form (nested) bubbles.

At four loops,
this is no longer the case 
and leads to a modification of the 
relation in \eqref{couprel}.
Divergences no longer cancel at the same loop-level, but the four-loop contribution to the anomalous dimension 
can be cancelled by a one-loop contribution 
if \eqref{couprel} is modified by a fourth-order term.
This was previously worked out for the (complex) $\beta$-deformation in
 \cite{Elmetti:2006gr,Elmetti:2007up} 
and for the entire Leigh-Strassler deformation \eqref{superpot}
in \cite{Bork:2007bj}. Here, we recall these results.

The only four-loop diagram without a bubble structure 
is given by
\begin{equation}\label{sefour}
\begin{aligned}
\csefour[\fmfi{plain}{p3}\fmfi{plain}{p4}\fmfi{plain}{vm1--vm2}]{plain}{plain}{plain}{plain}{plain}{plain}{plain}{plain}
&\to-N^4\frac{(\kappa\bar\kappa)^4}{2}((1+q\bar q+h\bar h)^4
+\delta H(q,\bar q,h,\bar h))I_{4\text{t}}
\col\\
\end{aligned}
\end{equation}
where the first term is 
is cancelled by the 
remaining four-loop diagrams. 
The additional 
contribution is a straightforward generalization of the 
result at $h=0$ in \cite{Elmetti:2006gr,Elmetti:2007up} and 
in accord with \cite{Bork:2007bj} we find
\begin{equation}\label{deltaH}
\begin{aligned}
\delta H(q,\bar q,h,\bar h)&=
(1-q\bar q-h\bar h)^4
+8q\bar qh\bar h(q\bar q(3-q\bar q)+h\bar h(3-h\bar h))\\
&\phantom{{}={}}
+8q\bar q(h^3+\bar h^3)
-8h\bar h(q^3+\bar q^3)
-8(q^3\bar h^3+\bar q^3h^3)
\pnt
\end{aligned}
\end{equation}
Since
this term is not cancelled by the other four-loop contributions
it  
will 
generate an anomalous dimension for the chiral field, 
unless one allows for fourth order corrections to 
 the marginal couplings. 
Unlike in 
the complex $\beta$-deformed case, it 
is not sufficient to have a series expansion 
 with constant coefficients
for the correction in \eqref{couprel}
Instead, we should 
make the ansatz
\begin{equation}\label{kmod}
\kappa\to\kappa(1+\lambda^3\Delta\kappa(q,\bar q,h,\bar h))\col\qquad
\bar\kappa\to\bar\kappa(1+\lambda^3\Delta\bar\kappa(q,\bar q,h,\bar h))
\pnt
\end{equation}
If we absorb these corrections into the 
original $\kappa$, $\bar\kappa$
the relation \eqref{couprel} is modified to 
\begin{equation}
2g_\YM^2=\kappa\bar\kappa(1+q\bar q+h\bar h)
(1-\lambda^3(\Delta\kappa+\Delta\bar\kappa))
\col
\end{equation}
where further modifications will occur at higher orders.
With these corrections, the leftover piece \eqref{deltaH} from 
the four-loop diagrams can be cancelled by the chiral 
one-loop diagram \eqref{sediag} with modified couplings \eqref{kmod}.
For conformal invariance we have to cancel the anomalous dimensions
which is achieved by setting
\begin{equation}
\begin{aligned}\label{gamma4cancel}
2
\bigg[\seone[\fmfcmd{fill fullcircle scaled 9 shifted vloc(__vc1) withcolor black ;}
\fmfiv{label=\small$\textcolor{white}{\scriptstyle\Delta}$,l.dist=0}{vloc(__vc1)}]{1}{1}{plain}{plain}{plain}{plain}{plain}{plain}
+
\seone[\fmfcmd{fill fullcircle scaled 9 shifted vloc(__vc2) withcolor black ;}
\fmfiv{label=\small$\textcolor{white}{\scriptstyle\Delta}$,l.dist=0}{vloc(__vc2)}]{1}{1}{plain}{plain}{plain}{plain}{plain}{plain}
\bigg]
+
8\bigg[
\csefour[\fmfi{plain}{p3}\fmfi{plain}{p4}\fmfi{plain}{vm1--vm2}]{plain}{plain}{plain}{plain}{plain}{plain}{plain}{plain}+\dots\bigg]
=\text{finite}
\col
\end{aligned}
\end{equation}
where the blobs are the vertex corrections which come with factors of $\lambda^3\Delta\kappa$ or $\lambda^3\Delta\bar\kappa$.
The ellipsis denotes the remaining four-loop 
diagrams in which the chiral field lines form (nested) bubbles.
The integer prefactors come from the definition 
of the dilatation operator (the anomalous dimension) in 
\eqref{DinZ}. This leads to a  multiplication of the $K$-loop contribution 
by a factor $2K$. 
From the condition \eqref{gamma4cancel} we find
\begin{equation}
\begin{aligned}
\Delta\kappa+\Delta\bar\kappa
=16\frac{\delta H(q,\bar q,h,\bar h)}{(1+q\bar q+h\bar h)^4}\frac{\mathcal{I}_{4\text{t}}}{\mathcal{I}_1}
\pnt
\end{aligned}
\end{equation}
With this adjustment the theory is conformal but 
not finite, 
since the diagrams that contribute to the chiral field renormalization do not have  the prefactors in \eqref{gamma4cancel} and hence do not cancel.



The alterations induced by \eqref{sefour} also affect the operator renormalization. At four-loop order this only concerns
the diagrams associated with the simplest chiral function $\chi(1)$. 
But 
the relation in \eqref{gamma4cancel} guarantees that 
these diagrams  obey 
\begin{equation}\label{chionefour}
\begin{aligned}
2\left[\settoheight{\eqoff}{$\times$}%
\setlength{\eqoff}{0.5\eqoff}%
\addtolength{\eqoff}{-13\unitlength}%
\raisebox{\eqoff}{%
\fmfframe(-1.5,3)(-11.5,3){%
\begin{fmfchar*}(20,20)
\chionei{plain}{plain}{plain}{plain}{plain}
\fmfcmd{fill fullcircle scaled 9 shifted vloc(__vc1) withcolor black ;}
\fmfiv{label=\small$\textcolor{white}{\scriptstyle\Delta}$,l.dist=0}{vloc(__vc1)}
\end{fmfchar*}}}
+
\settoheight{\eqoff}{$\times$}%
\setlength{\eqoff}{0.5\eqoff}%
\addtolength{\eqoff}{-13\unitlength}%
\raisebox{\eqoff}{%
\fmfframe(-1.5,3)(-11.5,3){%
\begin{fmfchar*}(20,20)
\chionei{plain}{plain}{plain}{plain}{plain}
\fmfcmd{fill fullcircle scaled 9 shifted vloc(__vc2) withcolor black ;}
\fmfiv{label=\small$\textcolor{white}{\scriptstyle\Delta}$,l.dist=0}{vloc(__vc2)}
\end{fmfchar*}}}
\right]
+8
\left[
\settoheight{\eqoff}{$\times$}%
\setlength{\eqoff}{0.5\eqoff}%
\addtolength{\eqoff}{-13\unitlength}%
\raisebox{\eqoff}{%
\fmfframe(-1.5,3)(-11.5,3){%
\begin{fmfchar*}(20,20)
\chionei{plain}{plain}{plain}{plain}{plain}
\fmfi{plain}{vd4--vd5}
\fmfi{plain}{vu4--vu5}
\fmfiset{p6}{vd4--vd5}
\fmfiset{p7}{vu4--vu5}
\svertex{vm6}{p6}
\svertex{vm7}{p7}
\fmfi{plain}{vm6--vm7}
\end{fmfchar*}}}
+\dots
\right]
=\text{finite}
\,, 
\end{aligned}
\end{equation}
whose validity is obvious if one realizes that the chiral operator (represented by the bold line) can be treated like a chiral vertex.  Hence, the diagrams in \eqref{chionefour} have the same topology as those in \eqref{gamma4cancel} and so have the same cancellation of  UV divergences.  These diagrams can then be omitted from the very beginning since they will not contribute to the dilatation operator.

\subsection{Diagrams with reducible chiral functions}

\label{subsec:nnmrange}

In the $SU(2)$ subsector of the (complex) $\beta$-deformed 
theories,
not all of the chiral functions 
in \eqref{chifuncdef} are independent.  Many chiral functions contain terms that antisymmetrize three neighboring sites.  Such terms are zero when there are only two flavors, leading to relations between different chiral functions.  
The simplest example is 
\begin{equation}\label{chiidentities1}
\chiop(1,2,1)|_{SU(2)}=\chiop(2,1,2)|_{SU(2)}=\rho^4q\bar q\chiop(1)|_{SU(2)}\col
\end{equation}
where $|_{SU(2)}$ indicates the projection onto the subsector.
The general set of relations 
are worked out in  appendix \ref{app:chifuncrel}. 
 The chiral functions $\chiop(1,2,1)$ and $\chiop(2,1,2)$ 
  first appear in the three-loop dilatation operator $\mathcal{D}_3$. 
The respective terms in \eqref{D1D2D3} arise from the chiral Feynman diagram
\begin{equation}
\begin{aligned}\label{crange3}
\settoheight{\eqoff}{$\times$}%
\setlength{\eqoff}{0.5\eqoff}%
\addtolength{\eqoff}{-12\unitlength}%
\raisebox{\eqoff}{%
\fmfframe(-0.5,2)(-5.5,2){%
\begin{fmfchar*}(20,20)
\chionetwoone
\end{fmfchar*}}}
&=\lambda^3I_3\chiop(1,2,1)
\col
\end{aligned}
\end{equation}
and its reflection. 
At four loops, the relation \eqref{chiidentities1} generalizes
to 
chiral functions with four arguments. 
Here there are two types of relations,  given by
\begin{equation}\label{chiidentities2}
\begin{gathered}
\begin{aligned}
\chiop(1,2,1,2)|_{SU(2)}=\chiop(1,2,3,2)|_{SU(2)}=\chiop(2,1,2,3)|_{SU(2)}&=\rho^4q\bar q\chiop(1,2)\col\\
\chiop(2,1,2,1)|_{SU(2)}=\chiop(3,2,1,2)|_{SU(2)}=\chiop(2,3,2,1)|_{SU(2)}&=\rho^4q\bar q\chiop(2,1)\col\\
\end{aligned}\\
\left.
\begin{aligned}
\chiop(1,2,1,3)|_{SU(2)}=\chiop(1,3,2,3)|_{SU(2)}\\
\chiop(3,2,3,1)|_{SU(2)}=\chiop(3,1,2,1)|_{SU(2)}
\end{aligned}
\right\}
=\rho^4q\bar q\chiop(1,3)\col\\
\end{gathered}
\end{equation}
where a relation and its reflection are displayed in tandem.
We call chiral functions ``reducible'' if they simplify
as in \eqref{chiidentities1} and \eqref{chiidentities2}
when projected onto the $SU(2)$ subsector.

In analogy to the three-loop diagram \eqref{crange3},  four-loop
 diagrams that come with chiral functions having 
 four arguments are chiral. 
Hence, each one is generated by one diagram.
After $\D$-algebra, we find
\begin{equation}
\begin{gathered}\label{crange4}
\settoheight{\eqoff}{$\times$}%
\setlength{\eqoff}{0.5\eqoff}%
\addtolength{\eqoff}{-12\unitlength}%
\raisebox{\eqoff}{%
\fmfframe(-1.5,2)(-6.5,2){%
\begin{fmfchar*}(20,20)
\chionetwoonetwoone
\end{fmfchar*}}}
=\lambda^4I_4\chiop(1,2,1,2)
\col\,\,
\settoheight{\eqoff}{$\times$}%
\setlength{\eqoff}{0.5\eqoff}%
\addtolength{\eqoff}{-12\unitlength}%
\raisebox{\eqoff}{%
\fmfframe(-1.5,2)(-1.5,2){%
\begin{fmfchar*}(20,20)
\chionetwothreetwo
\end{fmfchar*}}}
=\lambda^4I_{4\text{w}}\chiop(1,2,3,2)
\col\,\,
\settoheight{\eqoff}{$\times$}%
\setlength{\eqoff}{0.5\eqoff}%
\addtolength{\eqoff}{-12\unitlength}%
\raisebox{\eqoff}{%
\fmfframe(-0.5,2)(-0.5,2){%
\begin{fmfchar*}(20,20)
\chitwoonetwothree
\end{fmfchar*}}}
=\lambda^4I_4\chiop(2,1,2,3)
\col\\
\settoheight{\eqoff}{$\times$}%
\setlength{\eqoff}{0.5\eqoff}%
\addtolength{\eqoff}{-12\unitlength}%
\raisebox{\eqoff}{%
\fmfframe(-1.5,2)(-1.5,2){%
\begin{fmfchar*}(20,20)
\chionetwoonethree
\end{fmfchar*}}}
=\lambda^4I_{4\text{bb}}\chiop(1,2,1,3)
\col\,\,
\settoheight{\eqoff}{$\times$}%
\setlength{\eqoff}{0.5\eqoff}%
\addtolength{\eqoff}{-12\unitlength}%
\raisebox{\eqoff}{%
\fmfframe(-1.5,2)(-1.5,2){%
\begin{fmfchar*}(20,20)
\chionethreetwothree
\end{fmfchar*}}}
=\lambda^4I_{4\text{w}}\chiop(1,3,2,3)
\col\,\,
\end{gathered}
\end{equation}
as well as  analogous results for their reflections. 

There are also four-loop diagrams   having chiral function $\chiop(1,2,1)$
or $\chiop(2,1,2)$. 
These diagrams are constructed  by attaching  a vector propagator to  \eqref{crange3} or its reflection.
%
%
%
%
Since the one-loop chiral self energy is identically zero, 
the ends of the vector propagator much attach to different chiral propagators.
Moreover, there are further restrictions coming from the finiteness 
conditions of \cite{Sieg:2010tz}.
First, 
the one  chiral vertex in   \eqref{crange3} which is not part of a loop 
must remain out of any loop after adding the vector field interaction.  
Second, 
the vector propagator cannot attach to  the neighboring field lines  of  \eqref{crange3}, keeping the range of the diagram to three sites. 
The  diagrams that fulfill these constraints, along with their values,  are 
\begin{equation}
\begin{aligned}\label{chi121r3}
&
\settoheight{\eqoff}{$\times$}%
\setlength{\eqoff}{0.5\eqoff}%
\addtolength{\eqoff}{-12\unitlength}%
\raisebox{\eqoff}{%
\fmfframe(-0.5,2)(-5.5,2){%
\begin{fmfchar*}(20,20)
\chionetwoone
\fmfi{photon}{vm3{dir 180}..{dir 0}vm4}
\end{fmfchar*}}}
\col
\settoheight{\eqoff}{$\times$}%
\setlength{\eqoff}{0.5\eqoff}%
\addtolength{\eqoff}{-12\unitlength}%
\raisebox{\eqoff}{%
\fmfframe(-0.5,2)(-5.5,2){%
\begin{fmfchar*}(20,20)
\chionetwoone
\fmfi{photon}{vm3{dir 0}..{dir -90}vm5}
\end{fmfchar*}}}
\col
\settoheight{\eqoff}{$\times$}%
\setlength{\eqoff}{0.5\eqoff}%
\addtolength{\eqoff}{-12\unitlength}%
\raisebox{\eqoff}{%
\fmfframe(-0.5,2)(-5.5,2){%
\begin{fmfchar*}(20,20)
\chionetwoone
\fmfi{photon}{vm5{dir 90}..{dir 0}vm6}
\end{fmfchar*}}}
\col
\settoheight{\eqoff}{$\times$}%
\setlength{\eqoff}{0.5\eqoff}%
\addtolength{\eqoff}{-12\unitlength}%
\raisebox{\eqoff}{%
\fmfframe(-0.5,2)(-5.5,2){%
\begin{fmfchar*}(20,20)
\chionetwoone
\fmfi{photon}{vm6{dir -90}..{dir 180}vm7}
\end{fmfchar*}}}
\col
\settoheight{\eqoff}{$\times$}%
\setlength{\eqoff}{0.5\eqoff}%
\addtolength{\eqoff}{-12\unitlength}%
\raisebox{\eqoff}{%
\fmfframe(-0.5,2)(-5.5,2){%
\begin{fmfchar*}(20,20)
\chionetwoone
\fmfi{photon}{vm4{dir 180}..{dir 0}vm10}
\end{fmfchar*}}}
\col
\settoheight{\eqoff}{$\times$}%
\setlength{\eqoff}{0.5\eqoff}%
\addtolength{\eqoff}{-12\unitlength}%
\raisebox{\eqoff}{%
\fmfframe(-0.5,2)(-5.5,2){%
\begin{fmfchar*}(20,20)
\chionetwoone
\fmfi{photon}{vm7{dir 0}..{dir -90}vm9}
\end{fmfchar*}}}
\col
\settoheight{\eqoff}{$\times$}%
\setlength{\eqoff}{0.5\eqoff}%
\addtolength{\eqoff}{-12\unitlength}%
\raisebox{\eqoff}{%
\fmfframe(-0.5,2)(-5.5,2){%
\begin{fmfchar*}(20,20)
\chionetwoone
\fmfi{photon}{vm10{dir 180}..{dir -90}vm11}
\end{fmfchar*}}}
\col
\settoheight{\eqoff}{$\times$}%
\setlength{\eqoff}{0.5\eqoff}%
\addtolength{\eqoff}{-12\unitlength}%
\raisebox{\eqoff}{%
\fmfframe(-0.5,2)(-5.5,2){%
\begin{fmfchar*}(20,20)
\chionetwoone
\fmfi{photon}{vm10{dir 0}..{dir -90}vm12}
\end{fmfchar*}}}
\to-I_4
\col\\
&
\settoheight{\eqoff}{$\times$}%
\setlength{\eqoff}{0.5\eqoff}%
\addtolength{\eqoff}{-12\unitlength}%
\raisebox{\eqoff}{%
\fmfframe(-0.5,2)(-5.5,2){%
\begin{fmfchar*}(20,20)
\chionetwoone
\fmfi{photon}{vm4{dir 0}..{dir -90}vm8}
\end{fmfchar*}}}
\col
\settoheight{\eqoff}{$\times$}%
\setlength{\eqoff}{0.5\eqoff}%
\addtolength{\eqoff}{-12\unitlength}%
\raisebox{\eqoff}{%
\fmfframe(-0.5,2)(-5.5,2){%
\begin{fmfchar*}(20,20)
\chionetwoone
\fmfi{photon}{vm8{dir 90}..{dir 0}vm7}
\end{fmfchar*}}}
\to-I_{4\text{w}}
\col\qquad
\settoheight{\eqoff}{$\times$}%
\setlength{\eqoff}{0.5\eqoff}%
\addtolength{\eqoff}{-12\unitlength}%
\raisebox{\eqoff}{%
\fmfframe(-0.5,2)(-5.5,2){%
\begin{fmfchar*}(20,20)
\chionetwoone
\fmfi{photon}{vm8{dir -90}..{dir 180}vm10}
\end{fmfchar*}}}
\col
\settoheight{\eqoff}{$\times$}%
\setlength{\eqoff}{0.5\eqoff}%
\addtolength{\eqoff}{-12\unitlength}%
\raisebox{\eqoff}{%
\fmfframe(-0.5,2)(-5.5,2){%
\begin{fmfchar*}(20,20)
\chionetwoone
\fmfi{photon}{vm9{dir 180}..{dir 90}vm8}
\end{fmfchar*}}}
\to-I_{4\text{bt}}
\col\qquad
\settoheight{\eqoff}{$\times$}%
\setlength{\eqoff}{0.5\eqoff}%
\addtolength{\eqoff}{-12\unitlength}%
\raisebox{\eqoff}{%
\fmfframe(-0.5,2)(-5.5,2){%
\begin{fmfchar*}(20,20)
\chionetwoone
\fmfi{photon}{vm11{dir 0}..{dir 0}vm12}
\end{fmfchar*}}}
\to-I_{43\text{t}}
\col\\
&
\settoheight{\eqoff}{$\times$}%
\setlength{\eqoff}{0.5\eqoff}%
\addtolength{\eqoff}{-12\unitlength}%
\raisebox{\eqoff}{%
\fmfframe(-0.5,2)(-5.5,2){%
\begin{fmfchar*}(20,20)
\chionetwoone
\fmfi{photon}{vm3{dir 0}..{dir 0}vm6}
\end{fmfchar*}}}
\col
\settoheight{\eqoff}{$\times$}%
\setlength{\eqoff}{0.5\eqoff}%
\addtolength{\eqoff}{-12\unitlength}%
\raisebox{\eqoff}{%
\fmfframe(-0.5,2)(-5.5,2){%
\begin{fmfchar*}(20,20)
\chionetwoone
\fmfi{photon}{vm3{dir 180}..{dir 0}vm10}
\end{fmfchar*}}}
\col
\settoheight{\eqoff}{$\times$}%
\setlength{\eqoff}{0.5\eqoff}%
\addtolength{\eqoff}{-12\unitlength}%
\raisebox{\eqoff}{%
\fmfframe(-0.5,2)(-5.5,2){%
\begin{fmfchar*}(20,20)
\chionetwoone
\fmfi{photon}{vm6{dir -90}..{dir -90}vm9}
\end{fmfchar*}}}
\col
\settoheight{\eqoff}{$\times$}%
\setlength{\eqoff}{0.5\eqoff}%
\addtolength{\eqoff}{-12\unitlength}%
\raisebox{\eqoff}{%
\fmfframe(-0.5,2)(-5.5,2){%
\begin{fmfchar*}(20,20)
\chionetwoone
\fmfi{photon}{vm4{dir 0}..{dir 0}vm7}
\end{fmfchar*}}}
\col
\settoheight{\eqoff}{$\times$}%
\setlength{\eqoff}{0.5\eqoff}%
\addtolength{\eqoff}{-12\unitlength}%
\raisebox{\eqoff}{%
\fmfframe(-0.5,2)(-5.5,2){%
\begin{fmfchar*}(20,20)
\chionetwoone
\fmfi{photon}{vm10{dir 0}..{dir 0}vm9}
\end{fmfchar*}}}
\to I_4
\col\qquad
\settoheight{\eqoff}{$\times$}%
\setlength{\eqoff}{0.5\eqoff}%
\addtolength{\eqoff}{-12\unitlength}%
\raisebox{\eqoff}{%
\fmfframe(-0.5,2)(-5.5,2){%
\begin{fmfchar*}(20,20)
\chionetwoone
\fmfi{photon}{vm8{dir 90}..{dir 90}vm5}
\end{fmfchar*}}}
\to I_{4\text{bt}}
\col\\
&
\settoheight{\eqoff}{$\times$}%
\setlength{\eqoff}{0.5\eqoff}%
\addtolength{\eqoff}{-12\unitlength}%
\raisebox{\eqoff}{%
\fmfframe(-0.5,2)(-5.5,2){%
\begin{fmfchar*}(20,20)
\chionetwoone
\fmfi{photon}{vm4{dir 180}..{dir -90}vm11}
\end{fmfchar*}}}
\to I_4+I_{4\beta}-I_{4\text{tr}1}
\col\qquad
\settoheight{\eqoff}{$\times$}%
\setlength{\eqoff}{0.5\eqoff}%
\addtolength{\eqoff}{-12\unitlength}%
\raisebox{\eqoff}{%
\fmfframe(-0.5,2)(-5.5,2){%
\begin{fmfchar*}(20,20)
\chionetwoone
\fmfi{photon}{vm8{dir -90}..{dir -90}vm12}
\end{fmfchar*}}}
\to I_4+I_{4\text{w}}-I_{4\text{tr}2}
\col\\
&
\settoheight{\eqoff}{$\times$}%
\setlength{\eqoff}{0.5\eqoff}%
\addtolength{\eqoff}{-12\unitlength}%
\raisebox{\eqoff}{%
\fmfframe(-0.5,2)(-5.5,2){%
\begin{fmfchar*}(20,20)
\chionetwoone
\fmfi{photon}{vm3{dir 180}..{dir -90}vm11}
\end{fmfchar*}}}
\to-I_{4\text{w}}
\col\qquad
\settoheight{\eqoff}{$\times$}%
\setlength{\eqoff}{0.5\eqoff}%
\addtolength{\eqoff}{-12\unitlength}%
\raisebox{\eqoff}{%
\fmfframe(-0.5,2)(-5.5,2){%
\begin{fmfchar*}(20,20)
\chionetwoone
\fmfi{photon}{vm12{dir 45}..{dir 45}vm9}
\end{fmfchar*}}}
\to-I_4
\col
\end{aligned}
\end{equation}
where we have grouped together the diagrams that
lead to the same integrals  after $\D$-algebra.  The integrals are listed in \eqref{IK}.
The above results contain all signs from color and flavor factors, and from 
the $\D$-algebra manipulations. 
Finite contributions and a common factor
$\lambda^4\chiop(1,2,1)$ have been omitted.
Note that most of the diagrams 
are easily evaluated 
using the arguments
in \cite{Sieg:2010tz}. 
The diagrams in the fourth line
require a little more work and lead to integrals with momenta
in the numerator of their integrands that are contracted as prescribed
by a trace over products of $\gamma$-matrices.
Using the results of appendix \ref{app:integrals}, 
we find for the sum of the most complicated diagrams
\begin{equation}
\begin{aligned}
\settoheight{\eqoff}{$\times$}%
\setlength{\eqoff}{0.5\eqoff}%
\addtolength{\eqoff}{-12\unitlength}%
\raisebox{\eqoff}{%
\fmfframe(-0.5,2)(-5.5,2){%
\begin{fmfchar*}(20,20)
\chionetwoone
\fmfi{photon}{vm4{dir 180}..{dir -90}vm11}
\end{fmfchar*}}}
+
\settoheight{\eqoff}{$\times$}%
\setlength{\eqoff}{0.5\eqoff}%
\addtolength{\eqoff}{-12\unitlength}%
\raisebox{\eqoff}{%
\fmfframe(-0.5,2)(-5.5,2){%
\begin{fmfchar*}(20,20)
\chionetwoone
\fmfi{photon}{vm8{dir -90}..{dir -90}vm12}
\end{fmfchar*}}}
\to
-2I''_{4\text{t}1}
+I_{4\text{w}}
+I_{4\text{bt}}
+I_{43\text{t}}
-I_{4\text{t}}
\pnt\\
\end{aligned}
\end{equation}
Summing up all diagrams of \eqref{chi121r3}, and eliminating 
$I_{4\text{w}}$ by making use of the relation \eqref{I4I4wrel} 
results in
\begin{equation}\label{range3}
\begin{aligned}
\settoheight{\eqoff}{$\times$}%
\setlength{\eqoff}{0.5\eqoff}%
\addtolength{\eqoff}{-12\unitlength}%
\raisebox{\eqoff}{%
\fmfframe(-0.5,2)(-5.5,2){%
\begin{fmfchar*}(20,20)
\chionetwoone
\end{fmfchar*}}}
+\settoheight{\eqoff}{$\times$}%
\setlength{\eqoff}{0.5\eqoff}%
\addtolength{\eqoff}{-2\unitlength}%
\raisebox{\eqoff}{%
\fmfframe(1,0)(1,0){%
\begin{fmfchar*}(7.5,4)
\fmfleft{in}
\fmfright{out}
\fmf{photon}{in,out}
\end{fmfchar*}}}
=\lambda^4(-6I_4+2I''_{4\text{t}1}+I_{4\text{t}})\chiop(1,2,1)
\pnt
\end{aligned}
\end{equation}
The result for the reflected diagrams is obtained by 
replacing 
$\chiop(1,2,1)$ with $\chiop(2,1,2)$.

\subsection{Four-loop result}
\label{sec:fourloopresults}

The reducible contribution
 to the renormalization constant is the negative  sum of \eqref{range3},  \eqref{crange4} 
and  their reflections.  Using \eqref{DinZ}, one finds that the reducible part 
contributes to the dilatation operator with the coefficient of the $\frac{1}{\varepsilon}$ pole multiplied by $8$. 
Inserting the explicit expressions in \eqref{IK} then gives
\begin{equation}\label{deltaDred}
\begin{aligned}
\delta\mathcal{D}_{4\text{red}}
&=(60-8\zeta(3))[\chiop(1,2,1)+\chiop(2,1,2)]\\
&\phantom{{}={}}-10[\chiop(1,2,1,2)+\chiop(2,1,2,1)]\\
&
\phantom{{}={}}-(10-8\zeta(3))[\chiop(1,2,3,2)+\chiop(3,2,1,2)]\\
&
\phantom{{}={}}-10[\chiop(2,1,2,3)+\chiop(2,3,2,1)]\\
&
\phantom{{}={}}+\Big(\frac{10}{3}-4\zeta(3)\Big)[\chiop(1,2,1,3)+\chiop(3,2,3,1)]\\
&
\phantom{{}={}}-(10-8\zeta(3))[\chiop(1,3,2,3)+\chiop(3,1,2,1)]
\pnt
\end{aligned}
\end{equation}
Restricting to the $SU(2)$ subsector of  $\mathcal{N}=4$ SYM
 and using
 the identities in \eqref{chiidentities1} and
\eqref{chiidentities2},
reduces the above term to
\begin{equation}
\begin{aligned}\label{deltaDredSU2}
\delta\mathcal{D}_{4\text{red}}|_{SU(2)}
&=8(15-2\zeta(3))\chiop(1)
-2(15-4\zeta(3))[\chiop(1,2)+\chiop(2,1)]\\
&\phantom{{}={}}
-8\Big(\frac{5}{3}-\zeta(3)\Big)\chiop(1,3)
\pnt
\end{aligned}
\end{equation}

The four-loop dilatation operator $\DD_{4,\NN=4}$ for the $SU(2)$ subsector of $\NN=4$ SYM was first presented in \cite{Beisert:2007hz}. It is determined 
as one of the commuting charges of the integrable system, 
using information
from the all loop Bethe equations  \cite{Beisert:2005fw} and the magnon 
dispersion relation 
\cite{Santambrogio:2002sb,Beisert:2004hm,Beisert:2005tm}
\begin{equation}\label{Ep}
E(p)=\sqrt{1+4h^2(g)\sin^2\tfrac{p}{2}}-1\col
\end{equation}
where we assume that $h^2(g)=4g^2$.  To find $\DD_4$ for the Leigh-Strassler theories we add  \eqref{deltaDred} to $\DD_{4,\NN=4}$ and then subtract \eqref{deltaDredSU2}.  Using the convention in \cite{Fiamberti:2008sh} for $\DD_{4,\NN=4}$, we find
\begin{equation}
\begin{aligned}\label{D4chi}
\mathcal{D}_4&=
{}+{}16(5+\zeta(3))\chiop(1)\\
&\phantom{{}={}}
-8(15+\zeta(3))[\chiop(1,2)+\chiop(2,1)]
+8\Big(\frac{23}{3}-\zeta(3)\Big)\chiop(1,3)
-4\chiop(1,4)\\
&\phantom{{}={}}
+4(15-2\zeta(3))[\chiop(1,2,1)+\chiop(2,1,2)]
+60[\chiop(1,2,3)+\chiop(3,2,1)]\\
&\phantom{{}={}}
+2(4+\beta+2\epsilon_{3a}-2i\epsilon_{3b}+i\epsilon_{3c}-2i\epsilon_{3d})
\chiop(1,3,2)\\
&\phantom{{}={}}
+2(4+\beta+2\epsilon_{3a}+2i\epsilon_{3b}-i\epsilon_{3c}+2i\epsilon_{3d})
\chiop(2,1,3)\\
&\phantom{{}={}}
-2(2+2i\epsilon_{3b}+i\epsilon_{3c})[\chiop(1,2,4)+\chiop(1,4,3)]\\
&\phantom{{}={}}
-2(2-2i\epsilon_{3b}-i\epsilon_{3c})[\chiop(1,3,4)+\chiop(2,1,4)]\\
&\phantom{{}={}}
-10[\chiop(1,2,1,2)+\chiop(2,1,2,1)]\\
&\phantom{{}={}}
-(10+i\epsilon_{3e}-i\epsilon_{3f})[\chiop(2,1,2,3)+\chiop(2,3,2,1)]\\
&\phantom{{}={}}
-(10-8\zeta(3)-i\epsilon_{3e}+i\epsilon_{3f})[\chiop(1,2,3,2)+\chiop(3,2,1,2)]\\
&\phantom{{}={}}
+\Big(\frac{14}{3}+8\zeta(3)+2\epsilon_{3a}-4i\epsilon_{3b}+i\epsilon_{3e}\Big)[\chiop(1,3,2,3)+\chiop(3,1,2,1)]\\
&\phantom{{}={}}
+\Big(\frac{14}{3}-4\zeta(3)+2\epsilon_{3a}+4i\epsilon_{3b}-i\epsilon_{3e}\Big)[\chiop(1,2,1,3)+\chiop(3,2,3,1)]\\
&\phantom{{}={}}-2(6+\beta+2\epsilon_{3a})\chiop(2,1,3,2)\\
&\phantom{{}={}}
+2(9+2\epsilon_{3a})[\chiop(1,3,2,4)+\chiop(2,1,4,3)]\\
&\phantom{{}={}}
-2(4+\epsilon_{3a}+i\epsilon_{3b})[\chiop(1,2,4,3)+\chiop(1,4,3,2)]\\
&\phantom{{}={}}
-2(4+\epsilon_{3a}-i\epsilon_{3b})[\chiop(2,1,3,4)+\chiop(3,2,1,4)]\\
&\phantom{{}={}}
-10[\chiop(1,2,3,4)+\chiop(4,3,2,1)]\col
\end{aligned}
\end{equation}
where we have also considered the possibility of more general 
similarity transformations 
as
compared to 
$\mathcal{N}=4$ SYM.
These are parameterized by two additional parameters 
$\epsilon_{3e}$ and $\epsilon_{3f}$ 
and they occur because the identities in \eqref{chiidentities1} and  \eqref{chiidentities2}
are no longer applicable.
The details are worked out in appendix 
\ref{app:strafo}.
The dressing phase $\beta$ and the coefficients 
$\epsilon_{3a},\dots,\epsilon_{3c}$ of similarity transformations
in the scheme of $\mathcal{N}=1$ supergraphs are fixed by comparing
the $SU(2)$ subsector projection of the above result to the 
integrability-based expression.
This is not the case for $\epsilon_{3e}$ and $\epsilon_{3f}$
that drop out in the projection and hence can be set to convenient 
values. With our choice we have made the rational numbers
within the coefficients of $\chiop(1,3,2,3)+\chiop(3,1,2,1)$ and $\chiop(1,2,1,3)+\chiop(3,2,3,1)$
equal.
This yields
\begin{equation}
\beta=4\zeta(3)\col\qquad
\epsilon_{3a}=-4\col\qquad
\epsilon_{3b}=-i\frac{4}{3}\col\qquad
\epsilon_{3c}=i\frac{4}{3}\col\qquad
\epsilon_{3e}=i\frac{4}{3}\col\qquad
\epsilon_{3f}=i\frac{4}{3}
\col
\end{equation}
Note that we have not fixed 
the coefficient $\epsilon_{3d}$ in the 
scheme of Feynman diagrams in $\mathcal{N}=1$ superspace.
The calculation would be tedious and unnecessary here,
since the combination 
$\chiop(1,3,2)-\chiop(1,3,2)$ of chiral functions that are conjugate to each 
other vanishes whenever applied to the states, and hence 
$\epsilon_{3d}$ drops out.

\section{Application to single-impurity states}
\label{sec:oneimpstates}

We call chiral functions ``connected'' if adjacent 
entries in
their arguments 
differ 
by $\pm1$.
These are the only chiral functions that 
do not vanish 
when acting 
on the one-impurity states in \eqref{SU2oneexstate}
and \eqref{cubiconeexstate}. 
For this reason the magnon dispersion relations can only depend on the connected chiral functions.
Using 
the definition of a connected product of chiral functions 
in \eqref{connprod}, we can reexpress the the first four orders of $\DD$ in
\eqref{D1D2D3} and \eqref{D4chi} as
\begin{equation}\label{D1D2D3D4}
\begin{aligned}
\mathcal{D}_1&={}-{}2\chiop(1)\col\\
\mathcal{D}_2&={}-{}2\big[\chiop(1)^2\big]_{\text{c}}\col\\
\mathcal{D}_3&={}-{}4\big[\chiop(1)^3\big]_{\text{c}}
+4i\epsilon_2[\chiop(2,1,3)-\chiop(1,3,2)]-4\chiop(1,3)\col\\
\mathcal{D}_4&=
{}-{}10\big[\chiop(1)^4\big]_{\text{c}}\\
&\phantom{{}={}}
+8\zeta(3)[{}-{}2\chiop(1)+\chiop(1,2)+\chiop(2,1)\\
&\phantom{{}={}+8\zeta(3)[}
-\chiop(1,2,1)-\chiop(2,1,2)
+\chiop(1,2,3,2)+\chiop(3,2,1,2)]\\
&\phantom{{}={}}
-i(\epsilon_{3e}-\epsilon_{3f})[\chiop(2,1,2,3)+\chiop(2,3,2,1)-\chiop(1,2,3,2)-\chiop(3,2,1,2)]\\
&\phantom{{}={}}
+8\Big(\frac{23}{3}-\zeta(3)\Big)\chiop(1,3)
-4\chiop(1,4)\\
&\phantom{{}={}}
+2(4+\beta+2\epsilon_{3a}-2i\epsilon_{3b}+i\epsilon_{3c}-2i\epsilon_{3d})
\chiop(1,3,2)\\
&\phantom{{}={}}
+2(4+\beta+2\epsilon_{3a}+2i\epsilon_{3b}-i\epsilon_{3c}+2i\epsilon_{3d})
\chiop(2,1,3)\\
&\phantom{{}={}}
-2(2+2i\epsilon_{3b}+i\epsilon_{3c})[\chiop(1,2,4)+\chiop(1,4,3)]\\
&\phantom{{}={}}
-2(2-2i\epsilon_{3b}-i\epsilon_{3c})[\chiop(1,3,4)+\chiop(2,1,4)]\\
&\phantom{{}={}}
+\Big(\frac{14}{3}+8\zeta(3)+2\epsilon_{3a}-4i\epsilon_{3b}+i\epsilon_{3e}\Big)[\chiop(1,3,2,3)+\chiop(3,1,2,1)]\\
&\phantom{{}={}}
+\Big(\frac{14}{3}-4\zeta(3)+2\epsilon_{3a}+4i\epsilon_{3b}-i\epsilon_{3e}\Big)[\chiop(1,2,1,3)+\chiop(3,2,3,1)]\\
&\phantom{{}={}}-2(6+\beta+2\epsilon_{3a})\chiop(2,1,3,2)\\
&\phantom{{}={}}
+2(9+2\epsilon_{3a})[\chiop(1,3,2,4)+\chiop(2,1,4,3)]\\
&\phantom{{}={}}
-2(4+\epsilon_{3a}+i\epsilon_{3b})[\chiop(1,2,4,3)+\chiop(1,4,3,2)]\\
&\phantom{{}={}}
-2(4+\epsilon_{3a}-i\epsilon_{3b})[\chiop(2,1,3,4)+\chiop(3,2,1,4)]
\pnt
\end{aligned}
\end{equation}
The terms with connected products appear naturally in the expansion of the $\NN=4$  magnon dispersion relation \cite{Sieg:2010tz,Minahan:2011dd}.  In particular, the first term in $\DD_4$ is consistent with a conjecture made in
\cite{Minahan:2011dd} that the rational connected parts of $\DD_m$ are given by
\begin{equation}
\frac{\Gamma(3/2)}{\Gamma(m+1)\Gamma(3/2-m)}(-4)^m\big[\chi(1)^m\big]_{\text{c}}
\pnt
\end{equation} 
The conjecture is nontrivial since it allows us to separate terms that would have been equivalent in the $\beta$-deformed or undeformed theory.

The next term in $\DD_4$ is a sum of connected chiral functions multiplied by $8\zeta(3)$.  It vanishes when we project to the 
$SU(2)$ subsector in $\NN=4$ or the real $\beta$-deformed theory and hence
is consistent with the dispersion relation \eqref{Ep} with 
$h^2(g)=4g^2$. However, it contributes when two distinct types of 
scalar magnons approach each other, and hence is associated with 
the 
scattering matrix for different magnons.
In the complex $\beta$-deformed case it is also non-vanishing 
if a single magnon 
is within its interaction range and hence it starts to appear 
within the respective dispersion relation as a four-loop contribution to
$h^2(g)$ in \eqref{Ep}.
%

We can now use \eqref{D1D2D3D4} to determine the
four-loop anomalous dimensions for particular operators.

\subsection{Complex $\beta$-deformation}

The one-impurity operators in the complex $\beta$-deformations,  where $|q|\neq 1$ and $h=0$, are eigenstates of $\chi(1)$ with eigenvalue $c(q,\bar q)$, where
\begin{equation}
 c(q,\bar q)=\rho^2(1-q)(1-\bar q)=\frac{2(1-q)(1-\bar q)}{(1+q\bar q)}\col
\end{equation}
and we have eliminated $\rho$ by applying  \eqref{buildingblock}.
Furthermore, the connected products of $\chi(1)$ are products of $c(q,\bar  q)$.  Using the relation in \eqref{chifuncrel}, we find 
\begin{equation}
\gamma=\big[\sqrt{1+4g^2c(q,\bar q)}-1\big]-8\zeta(3)(1-q\bar q)^2c^2(q,\bar q)
+\mathcal{O}(g^{10})\pnt
\end{equation}
The part of $\gamma$ inside the square bracket comes from the connected products in  \eqref{D1D2D3D4}.  The transcendental term comes from those connected chiral functions that are not included in the connected products.  In the special case where $\beta$ is real,  the transcendental term drops out and 
we find $c(q,\bar q)=4\sin^2\pi\beta$. The remaining part of $\gamma$ is the energy coming from \eqref{Ep}
for one excitation with   momentum shifted to $p=2\pi\beta$ by the twisted boundary conditions.

\subsection{Cubic Leigh-Strassler deformation}

In the cubic Leigh-Strassler deformation, where $\rho\to0$, $\rho|h|=\sqrt{2}$,
the 3-string null operators are eigenstates of every chiral function.  For $\chi(1)$ the eigenvalue is $-2M$, where $M$ is the number of pairs of adjacent fields with the same flavor.  For all other chiral functions the eigenvalue is 0.  Hence, the eigenvalue of the connected product $[\chi(1)^m]_c$ is $(-2)^m M$.
Inserting these expressions into  \eqref{D1D2D3D4}, we find that the anomalous dimensions of these operators are $\gamma_M=M\gamma_1$, where
\begin{equation}\label{fourloopgamma}
\gamma_1=\big[\sqrt{1+8g^2}-1\big]-32\,\zeta(3)\,g^8+\mathcal{O}(g^{10})
\pnt
\end{equation}

Note that the four-loop result is consistent with a conjecture in \cite{Minahan:2011dd} for the rational part of $\gamma_1$, where it was proposed that it would have the  form in the square brackets to all orders in $g$, assuming that  cancellations similar to \eqref{chionefour} continue to hold\footnote{In  \cite{Minahan:2011dd} the cancellations were also shown to hold at five-loop order.}. However, there is also a transcendental contribution starting at four-loop order.  A natural way to view this is that instead of $g^2$, the square root depends on a $g$ dependent function $\tilde h^2(g)$, such that
\begin{equation}\label{gamma}
\gamma_1=\sqrt{1+8\tilde h^2(g)}-1\,,
\end{equation}
where
\begin{equation}
\tilde h^2(g)=g^2-8\zeta(3)\,g^8+\dots\,.
\end{equation}
Without this correction the strong coupling behavior of $\gamma_1$ would have been $\gamma_1\sim g$,  but with it $\gamma_1$  likely increases with a smaller power of $g$.

\section{Conclusions}
\label{sec:conclusions}

The main result of this paper is the construction of the scalar part of the four-loop chiral dilatation operator \eqref{D1D2D3D4}.  This construction is valid for any Leigh-Strassler deformation and it is complete for the closed scalar subsectors of these deformations.  Completeness for an arbitrary deformation would require the contributions to $\DD$ involving the gauginos $\psi_\alpha$.

For the closed subsectors we have explicitly found the four-loop anomalous dimensions for the one impurity operators in the complex $\beta$-deformed theory and for {\it every} 3-string null operator in the cubic Leigh-Strassler theory.  
These anomalous dimensions have $\zeta(3)$ terms that first appear at four loops\footnote{The transcendental contributions do come one loop after those found 
in \cite{Pomoni:2011jj} for the $\NN=2$ interpolating theory in 
\cite{Gadde:2010zi,Gadde:2009dj}.}.  
These transcendental terms show up because  they are present in the  
coefficients of connected chiral functions in $\DD_4$.  


From the perspective of $\NN=4$ SYM or its integrable deformations, the $\zeta(3)$ coefficients of the  connected (or disconnected) chiral functions must trace back to the BES dressing phase  \cite{Beisert:2006ez};
 the only place that transcendental terms appear in the all loop Bethe equations  \cite{Beisert:2005fw} is in the dressing phase and so it must be the source for these terms in the chiral dilatation operator.  However, in the complex $\beta$-deformed theory, these same terms are associated with the dispersion relation, since they contribute to the anomalous dimensions for single impurity operators.   This suggests  that the dispersion relation for the complex $\beta$-deformed theory, or the anomalous dimensions of the 3-string null operators in the cubic deformation,  are in principle derivable from the BES dressing phase. This assumes that the tuning mechanism as pictured in \eqref{gamma4cancel} and \eqref{chionefour} continues to hold at higher loops, otherwise further corrections could spoil this relation.


For the immediate future, one can use the disentangling of the chiral functions to compute next to leading order wrapping effects, extending the analysis 
in \cite{Sieg:2005kd,Fiamberti:2007rj,Fiamberti:2008sh,Fiamberti:2008sm,Fiamberti:2008sn}.  It would also be interesting to find the complete four-loop chiral dilatation operator.  For this it would be useful to find an extension of the chiral functions that could also include the gauginos.

\enlargethispage{\baselineskip}

\section*{Acknowledgements}
J.\ A.\ M.\ and C.\ S.\ thank the Perimeter Institute for kind hospitality.
J.\ A.\ M.\ also thanks the CTP at MIT for kind hospitality during the course of this
work. The work of J.\ A.\ M.\ is supported in part by Vetenskapr\aa{}det.
The work of C.\ S.\ is supported by DFG, SFB 647
\emph{Raum -- Zeit -- Materie. Analytische und Geometrische Strukturen}.

\appendix

\section{Relations between chiral functions}
\label{app:chifuncrel}

In the $SU(2)$ subsector, the
building block \eqref{buildingblock} of the chiral functions
\eqref{chifuncdef} reduces to
\begin{equation}
\Fop_{ij}|_{SU(2)}=\rho^2(\dperm-\unitmatrix)_{ij}\col
\end{equation}
where the deformed permutation and the identity on the r.h.s.\ are the ones
 given in \eqref{permdef} but
projected into the subspace of two flavors by applying \eqref{lambdaSU2proj}. 
Using the resulting expressions, one
can check that the r.h.s.\ obeys the relation
\begin{equation}\label{buildingblockid}
(\dperm-\unitmatrix)_{n\,n+1}(\dperm-\unitmatrix)_{n+1\,n+2}(\dperm-\unitmatrix)_{n\,n+1}
=q\bar q(\dperm-\unitmatrix)_{n\,n+1}
\pnt
\end{equation}
This relation has its origin in the fact that three adjacent elementary fields 
within the single trace of the composite operators of the $SU(2)$ subsector
carry maximally two different field flavors. The permutations
then fulfill the relation
\begin{equation}
\begin{aligned}
0
&=
\unitmatrix-\dperm_{n\,n+1}-\dperm_{n+1\,n+2}+\dperm_{n\,n+1}\dperm_{n\,n+1}+\dperm_{n+\,n+2}\dperm_{n\,n+1}
-\dperm_{n\,n+1}\dperm_{n+1\,n+2}\dperm_{n\,n+1}
\col
\end{aligned}
\end{equation}
where the r.h.s.\ at vanishing $\beta$-deformation $q=1$ 
is nothing else than the total antisymmetrizer 
$\epsilon_{i_ni_{n+1}i_{n+2}}\epsilon^{o_no_{n+1}o_{n+2}}$ between three field 
flavors when expressed in terms of permutations. In the undeformed case,
a respective relation was already worked out in  \cite{Beisert:2005wv}.
When inserting \eqref{buildingblockid} into the definition of the chiral 
functions \eqref{chifuncdef}, we find that in the 
$SU(2)$ subsector they are reducible as
\begin{equation}\label{chifuncrel}
\chiop(a_1,\dots,a_k,a,b,a,a_{k+4},\dots,a_n)|_{SU(2)}
=\rho^4q\bar q\chiop(a_1,\dots,a_k,a,a_{k+4},\dots,a_n)|_{SU(2)}
\pnt
\end{equation}

For the general Leigh-Strassler deformation, it is very easy to check
that the chiral functions 
fulfill
\begin{equation}\label{chifuncred}
\begin{aligned}
\chiop(a_1,\dots,a_k,a,a,a_{k+3},\dots,a_n)
&=-\rho^2(1+q\bar q+h\bar h)\chiop(a_1,\dots,a_k,a,a_{k+3},\dots,a_n)\\
&=-2\chiop(a_1,\dots,a_k,a,a_{k+3},\dots,a_n)+\mathcal{O}(\kappa^6)
\col
\end{aligned}
\end{equation}
where the second equality relies on \eqref{buildingblock}.
The previous relation between chiral functions is required in order 
to simplify the (noncommutative but 
associative) products of chiral functions. 
In order two define such products of chiral functions, we assume from 
now on that the length $L$ is always sufficiently 
large, i.e.\ $L\ge\kappa_a+\kappa_b-1$. Thereby, $\kappa_a$, $\kappa_b$
are the ranges, i.e.\ the numbers
of legs involved in the chiral interactions, 
of the two chiral functions that are to be multiplied.
In terms of the arguments of \eqref{chifuncdef} the range is defined as
\begin{equation}
\kappa_a=\max_{a_1,\dots,a_n}-\min_{a_1,\dots,a_n}+2\pnt
\end{equation}
We first introduce the commutator of two chiral functions that is given by
\begin{equation}\label{commdef}
\begin{aligned}
\comm{\chiop(a_1\dots,a_n)\,}{\,\chiop(b_1,\dots,b_p)}
&=\sum_{\min\limits_{b_1,\dots,b_p}-\max\limits_{a_1,\dots,a_n}-1}^{\max\limits_{b_1,\dots,b_p}-\min\limits_{a_1,\dots,a_n}+1}
\chiop(a_1+k,\dots,a_n+k,b_1,\dots,b_p)
\\
&\phantom{{}={}}-\sum_{\min\limits_{a_1,\dots,a_n}-\max\limits_{b_1,\dots,b_p}-1}^{\max\limits_{a_1,\dots,a_n}-\min\limits_{b_1,\dots,b_p}+1}
\chiop(b_1+k,\dots,b_p+k,a_1,\dots,a_n)
\pnt
\end{aligned}
\end{equation}
We call chiral functions ``connected'' if adjacent entries in their list of 
arguments never differ by more than $\pm 1$. For these chiral 
functions, we define the connected product as 
\begin{equation}\label{connprod}
[\chiop(a_1\dots,a_n)\chiop(b_1,\dots,b_p)]_{\text{c}}
=\sum_{k=b_1-a_n-1}^{b_1-a_n+1}\chiop(a_1+k,\dots,a_n+k,b_1,\dots,b_p)\col
\end{equation}
such that the result is again a connected chiral function.
It is understood that for multiple connected chiral functions 
appearing within $[\dots]_\text{c}$, the connected product is
applied for each pairwise multiplication.

\section{Similarity transformations}
\label{app:strafo}

The representation of the dilatation operator is not unique, but 
it may be transformed by a change of the basis of operators that 
does not alter its eigenvalues. In this appendix, we work out such
transformations.
We include non-unitary cases that allow us to remove the 
anti-Hermitean contributions in the three-loop dilatation 
operator given in \eqref{D1D2D3}.
The similarity transformations can be realized as
\begin{equation}\label{strafo}
\mathcal{D}'=\e^{-\chi}\mathcal{D}\e^{\chi}=\mathcal{D}+\delta\mathcal{D}
\col
\end{equation}
where $\chi$ is a linear combination of chiral functions.
We demand that the transformations preserve the structural constraints
coming from the underlying Feynman diagrams, i.e.\
at each loop order in the weak coupling expansion of 
$\mathcal{D}$ the transformation must not generate contributions
that involve chiral functions that can only appear at higher orders.
This is guaranteed if the weak coupling expansion of $\chi$ only contains 
those chiral functions that can be associated with Feynman diagrams at the 
considered order. In the $SU(2)$ subsector of the $\mathcal{N}=4$ SYM theory
the transformations are parameterized by one and four 
free parameters respectively in the three- and
four-loop contribution to the dilatation operator. 
Here, we have to abandon the relation \eqref{chiidentities1} and 
construct a more general transformation at four loops.
The ansatz is given by
\begin{equation}\label{stansatz}
\begin{aligned}
\chi&=g^2i\delta_{11}\chiop(1)
+g^4\Big(
i\delta_{21}\chiop(1)
+\frac{i}{2}\delta_{22}[\chiop(1,2)+\chiop(2,1)]\Big)\\
&\phantom{{}={}}
+g^6\Big(
i\delta_{31}\chiop(1)
+\frac{i}{2}\delta_{32}[\chiop(1,2)+\chiop(2,1)]
+i\delta_{33}\chiop(1,3)
+\frac{i}{2}\delta_{34}[\chiop(1,2,1)+\chiop(2,1,2)]\\
&\phantom{{}={}+g^6\Big(}
+\frac{1}{2}(i\delta_{35}+\delta_{36})\chiop(1,3,2)
+\frac{1}{2}(i\delta_{35}-\delta_{36})\chiop(2,1,3)\\
&\phantom{{}={}+g^6\Big(}
+\frac{i}{2}\delta_{37}[\chiop(1,2,3)+\chiop(3,2,1)]
\Big)
\col
\end{aligned}
\end{equation}
where we have considered the fact that the chiral functions of the 
first two contributions 
in the last row are adjoint to each other, but the loop integrals
of the respective diagrams are different.

Inserting this ansatz into \eqref{strafo} and expanding in powers of
$g$, we respectively 
obtain for the non-vanishing transformations at two and three loops
\begin{equation}
\begin{aligned}
\delta\mathcal{D}_3&=
-4i\epsilon_2[\chiop(1,3,2)-\chiop(2,1,3)]
\col\\
\delta\mathcal{D}_4&=
2(2\epsilon_{3a}-2i\epsilon_{3b}+i\epsilon_{3c}-2i\epsilon_{3d})
\chiop(1,3,2)
+2(2\epsilon_{3a}+2i\epsilon_{3b}-i\epsilon_{3c}+2i\epsilon_{3d})
\chiop(2,1,3)\\
&\phantom{{}={}}
-2i(2\epsilon_{3b}+\epsilon_{3c})[
\chiop(1,2,4)+\chiop(1,4,3)
-\chiop(1,3,4)-\chiop(2,1,4)]\\
&\phantom{{}={}}
-i(\epsilon_{3e}-\epsilon_{3f})[
\chiop(2,1,2,3)+\chiop(2,3,2,1)
-\chiop(1,2,3,2)-\chiop(3,2,1,2)]\\
&\phantom{{}={}}
+(2\epsilon_{3a}-4i\epsilon_{3b}+i\epsilon_{3e})[\chiop(1,3,2,3)+\chiop(3,1,2,1)]\\
&\phantom{{}={}}
+(2\epsilon_{3a}+4i\epsilon_{3b}-i\epsilon_{3e})[\chiop(1,2,1,3)+\chiop(3,2,3,1)]\\
&\phantom{{}={}}-4\epsilon_{3a}[\chiop(2,1,3,2)-\chiop(1,3,2,4)-\chiop(2,1,4,3)]\\
&\phantom{{}={}}
-2(\epsilon_{3a}+i\epsilon_{3b})[\chiop(1,2,4,3)+\chiop(1,4,3,2)]\\
&\phantom{{}={}}
-2(\epsilon_{3a}-i\epsilon_{3b})[\chiop(2,1,3,4)+\chiop(3,2,1,4)]
\col
\end{aligned}
\end{equation}
where the independent parameters read
\begin{equation}
\begin{aligned}
\epsilon_2&=\frac{1}{2}(2\delta_{11}-\delta_{22})\col\\
\epsilon_{3a}&=-\frac{3}{2}\delta_{11}(2\delta_{11}-\delta_{22})
-\frac{1}{2}\delta_{36}
\col\qquad
&\epsilon_{3b}&=2\delta_{11}-\frac{1}{2}(\delta_{35}+\delta_{37})
\col\\
\epsilon_{3c}&=-2\delta_{11}-\delta_{33}+\delta_{35}+\delta_{37}
\col\qquad
&\epsilon_{3d}&=\delta_{21}+\delta_{22}-12\delta_{11}-\frac{1}{2}\delta_{32}+\frac{3}{2}\delta_{35}+\delta_{37}
\col\\
\epsilon_{3e}&=\delta_{34}-3\delta_{35}-\delta_{37}
\col\qquad
&\epsilon_{3f}&=-3\delta_{35}
\pnt
\end{aligned}
\end{equation}

\newpage

\section{Integrals}
\label{app:integrals}
\enlargethispage{\baselineskip}

All four-loop integrals $I$ and their pole parts $\mathcal{I}$ used in the 
text are given by
\begin{equation}\label{IK}
\begin{aligned}
I_{4\text{bb}}&=
\intsixpo{%
\fmf{plain}{v1,v2}
\fmf{plain}{v2,v3}
\fmf{plain}{vc,v1}
\fmf{plain}{vc,v2}
\fmf{plain,left=0.25}{v3,v4}
\fmf{plain,right=0.25}{v3,v4}
\fmf{plain,left=0.25}{vc,v4}
\fmf{plain,right=0.25}{vc,v4}}
\col\qquad
\mathcal{I}_{4\text{bb}}
=\frac{1}{(4\pi)^8}\Big(-\frac{1}{12\varepsilon^4}+\frac{1}{4\varepsilon^3}-\frac{1}{12\varepsilon^2}+\frac{1}{\varepsilon}\Big(-\frac{5}{12}+\frac{1}{2}\zeta(3)\Big)\Big) 
\col\\
I_4&=
\intsixpo{%
\fmf{plain}{v1,v2}
\fmf{plain}{v2,v3}
\fmf{plain}{v3,v4}
\fmf{plain}{vc,v1}
\fmf{plain}{vc,v2}
\fmf{plain}{vc,v3}
\fmf{plain,left=0.25}{vc,v4}
\fmf{plain,right=0.25}{vc,v4}}
\col\qquad
\mathcal{I}_4
=\frac{1}{(4\pi)^8}\Big(-\frac{1}{24\varepsilon^4}+\frac{1}{4\varepsilon^3}-\frac{19}{24\varepsilon^2}+\frac{5}{4\varepsilon}\Big) 
\col\\
I_{4\text{c}}&=
\intsixpo{%
\fmf{plain}{v1,v2}
\fmf{plain}{v3,v4}
\fmf{plain}{vc,v1}
\fmf{plain}{vc,v2}
\fmf{plain}{vc,v3}
\fmf{plain,left=0.25}{v2,v3}
\fmf{plain,right=0.25}{v2,v3}
\fmf{plain}{vc,v4}}
\col\qquad
\mathcal{I}_{4\text{c}}
=\frac{1}{(4\pi)^8}\Big(-\frac{1}{12\varepsilon^4}+\frac{5}{12\varepsilon^3}-\frac{13}{12\varepsilon^2}+\frac{1}{\varepsilon}\Big(\frac{11}{12}-\frac{1}{2}\zeta(3)\Big)\Big)
\col\\
I_{4\text{w}}&=
\intsixp{%
\fmf{plain}{v1,v3}
\fmf{plain}{v3,v4}
\fmf{plain}{v4,v5}
\fmf{plain}{v5,v1}
\fmf{plain}{vc,v3}
\fmf{plain}{vc,v4}
\fmf{plain,left=0.25}{vc,v5}
\fmf{plain,right=0.25}{vc,v5}}
\col\qquad
\mathcal{I}_{4\text{w}}
=\frac{1}{(4\pi)^8}\Big(-\frac{1}{24\varepsilon^4}+\frac{1}{4\varepsilon^3}-\frac{19}{24\varepsilon^2}+\frac{1}{\varepsilon}\Big(\frac{5}{4}-\zeta(3)\Big)\Big) 
\col\\
I_{4\beta}&=
\intsixp{%
\fmf{plain}{v1,v3}
\fmf{plain}{v3,v4}
\fmf{plain}{v4,v5}
\fmf{plain}{v5,v1}
\fmf{plain}{vc,v3}
\fmf{plain}{vc,v5}
\fmf{plain,left=0.25}{vc,v4}
\fmf{plain,right=0.25}{vc,v4}}
\col\qquad
\mathcal{I}_{4\beta}
=\frac{1}{(4\pi)^8}\Big(
-\frac{1}{12\varepsilon^4}+\frac{1}{3\varepsilon^3}
-\frac{5}{12\varepsilon^2}
-\frac{1}{\varepsilon}\Big(\frac{1}{2}-\zeta(3)\Big)\Big)
\col\\
I_{4\text{bt}}&=
\intsixp{%
\fmf{plain}{v1,v3}
\fmf{plain}{v3,v4}
\fmf{plain}{v4,v5}
\fmf{plain,left=0.25}{v5,v1}
\fmf{plain,right=0.25}{v5,v1}
\fmf{plain}{vc,v3}
\fmf{plain}{vc,v4}
\fmf{plain}{vc,v5}}
\col\qquad
\mathcal{I}_{4\text{bt}}
=\frac{1}{(4\pi)^8}\Big(-\frac{1}{2\varepsilon^2}\zeta(3)+\frac{1}{\varepsilon}\Big(\frac{3}{2}\zeta(3)+\frac{\pi^4}{120}\Big)\Big)
\col\\
I_{43\text{t}}&=
\intsixp[29]{%
\fmf{plain}{v1,v3}
\fmf{plain}{v3,v4}
\fmf{plain}{v4,v5}
\fmf{plain}{v5,v1}
\fmf{plain}{v3,v5}
\fmfpoly{phantom}{v3,v5,v7}
\fmf{plain}{v3,v7}
\fmf{plain}{v4,v7}
\fmf{plain}{v5,v7}
}
\col\qquad
\mathcal{I}_{43\text{t}}
=\frac{1}{(4\pi)^8}\Big(-\frac{3}{2\varepsilon^2}\zeta(3)+\frac{1}{\varepsilon}\Big(\frac{1}{2}\zeta(3)-\frac{\pi^4}{120}\Big)\Big)
\col\\
I_{4\text{t}}&=
\fourlint{plain}{plain}{plain}{plain}{plain}{plain}{plain}{plain}
\col\qquad
\mathcal{I}_{4\text{t}}
=\frac{1}{(4\pi)^8}\frac{1}{\varepsilon}5\zeta(5)
\col\\
I''_{4\text{t}1}&=
\intsixp[18]{%
\fmf{plain}{v1,v2}
\fmf{plain}{v2,v3}
\fmf{phantom}{v5,v7}
\fmf{phantom}{v7,v6}
\fmffreeze
\fmf{derplain,right=0.5}{v7,v3}
\fmf{derplain}{v7,v1}
\fmf{plain}{v2,vi1}
\fmf{plain}{v7,vi1}
\fmf{plain}{v3,vi2}
\fmf{plain}{v7,vi2}
\fmffreeze
\fmf{plain}{vi1,vi2}}
\col\qquad
\mathcal{I}''_{4\text{t}1}
=\frac{1}{(4\pi)^8}\frac{1}{2\varepsilon}(\zeta(3)-5\zeta(5))
\col\\
I''_{4\text{t}2}&=
\intsixp[18]{%
\fmf{plain}{v1,v2}
\fmf{plain}{v2,v3}
\fmf{phantom}{v5,v7}
\fmf{phantom}{v7,v6}
\fmffreeze
\fmf{plain,right=0.5}{v7,v3}
\fmf{plain}{v7,v1}
\fmf{derplain}{v2,vi1}
\fmf{plain}{v7,vi1}
\fmf{derplain}{v3,vi2}
\fmf{plain}{v7,vi2}
\fmffreeze
\fmf{plain}{vi1,vi2}}
=
I''_{4\text{t}1}+\frac{1}{2}(I_4-I_{4\text{w}}-I_{43\text{t}}+I_{4\text{t}})
\col\\
I_{4\text{tr}1}&=
\intsixp{%
\fmf{plain}{v1,v2}
\fmf{plain,left=0.5}{v2,v4}
\fmf{derplain,label=$\scriptscriptstyle \delta$,l.side=left,l.dist=2}{v4,v5}
\fmf{plain}{v5,v6}
\fmf{plain}{v6,v1}
\fmf{derplain,label=$\scriptscriptstyle \alpha$,l.side=left,l.dist=2}{v6,vi1}
\fmf{plain}{vi1,v2}
\fmf{plain}{v5,vi2}
\fmf{phantom}{vi2,v3}
\fmf{derplain,label=$\scriptscriptstyle \beta$,l.side=left,l.dist=2}{vi1,vi2}
\fmffreeze
\fmf{derplain,label=$\scriptscriptstyle \gamma$,l.side=left,l.dist=2}{vi2,v4}
}\tr(\gamma^\alpha\gamma^\beta\gamma^\gamma\gamma^\delta)
\col\\
I_{4\text{tr}2}&=
\intsixp{%
\fmf{plain}{v1,v2}
\fmf{plain,left=0.5}{v2,v4}
\fmf{derplain,label=$\scriptscriptstyle \gamma$,l.side=left,l.dist=2}{v4,v5}
\fmf{derplain,label=$\scriptscriptstyle \beta$,l.side=left,l.dist=2}{v5,v6}
\fmf{plain}{v6,v1}
\fmf{derplain,label=$\scriptscriptstyle \alpha$,l.side=left,l.dist=2}{v6,vi1}
\fmf{plain}{vi1,v2}
\fmf{plain}{v5,vi2}
\fmf{phantom}{vi2,v3}
\fmf{plain}{vi1,vi2}
\fmffreeze
\fmf{derplain,label=$\scriptscriptstyle \delta$,l.side=left,l.dist=2}{vi2,v4}
}\tr(\gamma^\alpha\gamma^\beta\gamma^\gamma\gamma^\delta)
\pnt
\end{aligned}
\end{equation}
We only need the sum of the last two integrals. This is much easier to work out than
the individual integrals due to the symmetrization in pairs of indices that occurs 
after decomposing the numerator momentum with Lorentz index $\beta$ in terms of 
the other two momenta at a vertex. One obtains the relation 
\begin{equation}
\begin{aligned}
I_{4\text{tr}1}+I_{4\text{tr}1}
&=2I''_{4\text{t}2}
-I_{4\text{bt}}
+I_4
+I_{4\text{w}}
+I_{4\beta}
\pnt
\end{aligned}
\end{equation}
Note also that $\mathcal{I}_{4\text{w}}$ differs from $\mathcal{I}_4$ 
only by an additional $-\zeta(3)$ in the simple $\frac{1}{\varepsilon}$ pole.
Using then
\begin{equation}\label{zeta3simplepole}
2\mathcal{I}''_{4\text{t}1}
+\mathcal{I}_{4\text{t}}=\frac{1}{(4\pi)^8}\frac{1}{\varepsilon}\zeta(3)
\col
\end{equation}
we can express $\mathcal{I}_{4\text{w}}$ in terms of $\mathcal{I}_4$ as
\begin{equation}\label{I4I4wrel}
\mathcal{I}_{4\text{w}}=\mathcal{I}_4-2\mathcal{I}''_{4\text{t}1}
-\mathcal{I}_{4\text{t}}
\pnt
\end{equation}

\end{fmffile}


\footnotesize
\bibliographystyle{JHEP}
\bibliography{references}

\end{document}